\documentclass[11pt]{article}
\usepackage{fullpage}
\usepackage{times}

\usepackage{latexsym} 
\usepackage{amsmath} 
\usepackage{cite} 
\usepackage{multirow}
\usepackage[dvips]{graphics,color}

\newtheorem{theorem}{Theorem}[section]
\newtheorem{lemma}[theorem]{Lemma}

\newtheorem{corollary}[theorem]{Corollary}

\newenvironment{proof}{\noindent{\bf Proof:}}{\hfill\ensuremath{\Box}}

\newcommand{\ignore}[1]{}

\newcommand{\nca}[1]{\mbox{\it nca\ensuremath{(#1)}}}

\newcommand{\micro}[1]{\mbox{\it micro\ensuremath{(#1)}}}
\newcommand{\mroot}[1]{\mbox{\it root\ensuremath{(#1)}}}

\newcommand{\pair}[1]{\ensuremath{\{#1\}}}
\newcommand{\mlink}[1]{\mbox{\it link\ensuremath{(#1)}}}
\newcommand{\meval}[1]{\mbox{\it eval\ensuremath{(#1)}}}
\newcommand{\findroot}[1]{\mbox{\it findroot\ensuremath{(#1)}}}

\newcommand{\cut}[1]{\mbox{\it cut\ensuremath{(#1)}}}

\newcommand{\shp}[1]{\mbox{\it shp\ensuremath{(#1)}}}
\newcommand{\shc}[1]{\mbox{\it shc\ensuremath{(#1)}}}

\newcommand{\nul}{\mbox{\it null}}

\newcommand{\size}[1]{\mbox{\it size\ensuremath{(#1)}}}
\newcommand{\subsize}[1]{\mbox{\it subsize\ensuremath{(#1)}}}

\newcommand{\rank}[1]{\mbox{\it rank\ensuremath{(#1)}}}
\newcommand{\maxrank}[1]{\mbox{\it maxrank\ensuremath{(#1)}}}

\newcommand{\assign}[1]{\mbox{\it assign\ensuremath{(#1)}}}
\newcommand{\compress}[1]{\mbox{\it compress\ensuremath{(#1)}}}

\newcommand{\idom}[1]{\mbox{\it idom\ensuremath{(#1)}}}
\newcommand{\sdom}[1]{\mbox{\it sdom\ensuremath{(#1)}}}
\newcommand{\rdom}[1]{\mbox{\it rdom\ensuremath{(#1)}}}
\newcommand{\rd}[1]{\mbox{\it rd\ensuremath{(#1)}}}

\newcommand{\et}[1]{\mbox{\it et\ensuremath{(#1)}}}
\newcommand{\at}[1]{\mbox{\it at\ensuremath{(#1)}}}
\newcommand{\mt}[1]{\mbox{\it mt\ensuremath{(#1)}}}
\newcommand{\ct}[1]{\mbox{\it ct\ensuremath{(#1)}}}

\newcommand{\mv}[1]{\mbox{\it mid\ensuremath{(#1)}}}

\newcommand{\num}[1]{\mbox{\it num\ensuremath{(#1)}}}

\newcommand{\unite}{\mbox{\it unite}}
\newcommand{\find}{\mbox{\it find}}

\newcommand{\GG}{\ensuremath{{\cal G}}}

\newcommand{\boruvka}{Bor{\r u}vka}

\newcommand{\ltop}[1]{\mbox{\it top\ensuremath{(#1)}}}
\newcommand{\lbottom}[1]{\mbox{\it bottom\ensuremath{(#1)}}}

\newcommand{\anc}{\stackrel{{\scriptscriptstyle \ast}}{\rightarrow}}
\newcommand{\panc}{\stackrel{{\scriptscriptstyle +}}{\rightarrow}}

\newcommand{\oeval}{\mbox{\it eval}}
\newcommand{\ofindroot}{\mbox{\it findroot}}

\makeatletter
\def\argmin{\mathop{\operator@font argmin}}
\makeatother

\renewcommand{\eqref}[1]{Eq.~(\ref{#1})}

\title{
    Linear-Time Pointer-Machine Algorithms
    for Path-Evaluation Problems on Trees and Graphs\thanks{This work
	is partially covered by the extended abstracts,
	``Linear-Time Pointer-Machine Algorithms
	for Least Common Ancestors, MST Verification, and Dominators,''
        {\em Proc.~30th ACM Symp.~on Theory of Computing,}
        pp.~279--888, 1998,
        and ``Finding Dominators Revisited,''
        {\em Proc.~15th ACM-SIAM Symp.~on Discrete Algorithms,}
        pp.~862--871, 2004.
    }
}

\author{Adam L. Buchsbaum\thanks{
    AT\&T Labs--Research,
    Shannon Laboratory,
    180 Park Ave.,
    Florham Park, NJ 07932, USA;
    email: alb@research.att.com.}
  \and
    Loukas Georgiadis\thanks{
    Dept.~of Computer Science,
    University of Aarhus,
    IT-parken, Aabogade 34,
    DK-8200 Aarhus N, Denmark;
    email: loukas@daimi.au.dk.   
    Work partially done at Princeton University
    and partially supported by the National Science Foundation
    under the Aladdin Project, Grant No.~CCR-0122581.}
  \and
    Haim Kaplan\thanks{
    School of Mathematical Sciences,
    Tel Aviv University,
    Tel Aviv, Israel;
    email: haimk@math.tau.ac.il.
    Work partially done while a member of AT\&T Labs.}
  \and
    Anne Rogers\thanks{
    Dept.~of Computer Science,
    University of Chicago,
    1100 E 58th Street,
    Chicago, IL 60637, USA;
    email: amr@cs.uchicago.edu.
    Work partially done while a member of AT\&T Labs.}
  \and
    Robert E.~Tarjan\thanks{
    Dept.~of Computer Science,
    Princeton University,
    Princeton NJ, 08544, USA;
    and
    Hewlett-Packard, Palo Alto, CA;
    email: ret@cs.princeton.edu.
    Work at Princeton University partially supported by the National Science Foundation
    under the Aladdin Project, Grant No.~CCR-0122581.}
  \and
    Jeffery R.~Westbrook\thanks{
    Los Angeles, CA, USA;
    email: jwestbrook@acm.org.
    Work partially done while a member of AT\&T Labs.}
}

\date{October 31, 2006}

\begin{document}

\maketitle

\begin{abstract}
We present algorithms that run in linear time on pointer machines
for a collection of problems, each of which either directly or
indirectly requires the evaluation of a function defined on paths in
a tree.  These problems previously had linear-time algorithms but
only for  random-access machines (RAMs); the best pointer-machine
algorithms were super-linear by an inverse-Ackermann-function factor.
Our algorithms are also simpler, in some cases substantially, than
the previous linear-time RAM algorithms.  Our improvements come
primarily from three new ideas: a refined analysis of path
compression that gives a linear bound if the compressions favor
certain nodes, a pointer-based radix sort as a replacement for
table-based methods, and a more careful partitioning of a tree into
easily managed parts.  Our algorithms compute nearest common
ancestors off-line, verify and construct minimum spanning trees, do
interval analysis on a flowgraph, find the dominators of a
flowgraph, and build the component tree of a weighted tree.

\end{abstract}

\section{Introduction}
\label{sec:intro}

\begin{table}[t]
\caption{Time bounds.
$n$ is the number of vertices,
and
$m$ is either the number of edges/arcs for graph problems
or the number of NCA queries for the NCA problem.
$\alpha(m,n)$ is the standard functional inverse
of the Ackermann function.}
\label{tab:gaps}
\begin{center}
\begin{tabular}{l|ll|ll}
\hline
\multicolumn{1}{c|}{Problem} & \multicolumn{2}{c|}{Previous Pointer-Machine Bound} & \multicolumn{2}{c}{Previous RAM Bound} \\ \hline
Off-line NCAs &
    $O(m\alpha(m,n)+n)$ & \cite{lca:ahu} &
    $O(n+m)$ & \cite{nca:ht,lca:sv} \\
MST Verification &
    $O(m\alpha(m,n)+n)$ & \cite{pathcomp:t} &
    $O(n+m)$ & \cite{mst:drt:j,mstver:j:king} \\
MST Construction &
    $O(m\alpha(m,n)+n)$ & \cite{mst:c00} &
    $O(n+m)$ & \cite{mstj:fw,mst:kkt} \\
Interval Analysis &
    $O(m\alpha(m,n)+n)$ & \cite{st:t} &
    $O(n+m)$ & \cite{dsu:gt,st:t} \\
Dominators &
    $O(m\alpha(m,n)+n)$ & \cite{domin:lt} &
    $O(n+m)$ & \cite{domin:ahlt99,domin:bkrw} \\
Component Trees &
    $O(m\alpha(m,n)+n)$ &  &
    $O(n+m)$ & \cite{usp:t99} \\
    \hline
\end{tabular}
\end{center}
\end{table}

We study six problems---off-line computation of nearest common
ancestors (NCAs), verification and construction of minimum spanning
trees (MSTs), interval analysis of flowgraphs, finding dominators in
flowgraphs, and building the component tree of a weighted tree---that 
directly or indirectly require the evaluation of a function
defined on paths in a tree.  Each of these problems has a
linear-time algorithm on a RAM, but the fastest pointer-machine
algorithm is slower by an inverse-Ackermann-function factor.\footnote{
We use Tarjan's definition \cite{setunion:tarjan}.
Let $A(i,j)$ for $i,j \geq 1$ be defined by $A(1,j)=2^j$ for $j\geq 1$;
$A(i,1)=A(i-1,2)$ for $i\geq 2$;
and $A(i,j)=A(i-1,A(i,j-1))$ for $i,j\geq 2$.
Then $\alpha(m,n)=\min\{i\geq 1 : A(i,\lfloor m/n \rfloor) > \log n\}$.}
(See
Table~\ref{tab:gaps}.)  A pointer machine \cite{setunionptr:tarjan}
allows binary comparisons and arithmetic operations on data,
dereferencing of pointers, and equality tests on pointers.  It does
not permit pointer arithmetic or tests on pointers other than
testing for equality and is thus less powerful than the RAM model
\cite{algs:ahu}. Pointer machines are powerful enough to simulate
functional programming languages like LISP and ML.  Often,
though, the lack of random access complicates the design of
efficient pointer machine algorithms; the RAM algorithms for the
problems we consider rely on $O(1)$-time table lookup methods that
are not implementable on a pointer machine.  Nevertheless, we are
able to overcome the weaknesses of the pointer machine model and
develop linear-time algorithms for all six problems.  Not only are
our algorithms asymptotically as fast as the fastest RAM algorithms,
they are simpler too, in some cases substantially.

Our improvements come mainly from three new ideas.  The first is a
refined analysis of path compression.  Path compression is a
well-known technique first used to speed up the standard
disjoint-set-union (DSU) data structure \cite{setunion:tarjan} and
later extended to speed up the evaluation of functions defined on
paths in trees\cite{pathcomp:t}. Our applications use either the DSU
structure or path evaluation for the function \emph{minimum} or
\emph{maximum}, or both. We show that, under a certain restriction
on the compressions satisfied by our applications,
compression takes constant rather than
inverse-Ackermann amortized time.

The second new idea is to replace the table-based methods of the RAM
algorithms with a pointer-based radix sort.  Each of the RAM
algorithms precomputes answers to small subproblems, stores the
answers in a table, and looks up the answers by random access.  If
the size of the subproblems is small enough, the total size of all
distinct subproblems and the total time to solve them are linear
(or even sublinear) in the size of the original problem.  Our
alternative approach is to construct a pointer-based encoding of
each subproblem, group isomorphic subproblems together using a
pointer-based radix sort, solve one instance of each group of
isomorphic subproblems, and transfer its solution to the isomorphic
subproblems, all of which can be done on a pointer machine.

The third new idea is to change the partitioning strategy.  In order
to reduce the original problem to a collection of small subproblems,
the RAM algorithms partition a tree corresponding to the original
problem into small subtrees.  For some of the problems, partitioning
the entire tree into subtrees produces serious technical
complications; this is especially true of the dominators problem.
Instead, for all but one of the problems we partition only the
bottom part of the tree into small subtrees.  For NCAs and MSTs,
this together with our refined analysis of path compression 
suffices to yield a linear-time algorithm.  For interval analysis and
finding dominators, we also partition the remainder of the tree into
a set of maximal disjoint paths.  Only one of our applications,
building a component tree, relies on the original idea of
partitioning the entire tree into small subtrees.

The remainder of our paper proceeds as follows.  Section
\ref{sec:problems} formally defines the problems we consider and reviews
previous work. Section \ref{sec:pc} discusses disjoint set union,
computing path minima on trees, and a refined analysis of path
compression. Section \ref{sec:tgc} discusses the use of
pointer-based radix sorting to solve a graph problem for a
collection of many small instances.  Sections \ref{sec:nca} through
\ref{sec:ct} discuss our applications:  NCAs, MSTs, flowgraph
interval analysis, finding dominators, and building a component
tree, respectively. Section \ref{sec:remarks} contains concluding
remarks.  Our paper is a significantly revised and improved
combination of two conference papers\cite{ptrs:bkrw,dom:gt04},
including new results in Sections \ref{sec:ia} and \ref{sec:ct}.

\section{Problem Definitions and Previous Work}
\label{sec:problems}

Throughout this paper we denote the base-two logarithm by $\log$. We
assume $n \ge 2$ throughout.

\subsection{Nearest Common Ancestors}

\begin{description}
\item[Off-Line Nearest Common Ancestors:] Given an $n$-node tree $T$ rooted
at node $r$ and a set $P$ of $m$ node pairs, find, for each pair
$\pair{v,w}$ in $P$, the nearest common ancestor of $v$ and $w$ in
$T$, denoted by $\nca{v,w}$.
\end{description}

The fastest previous pointer-machine algorithm is that of Aho,
Hopcroft, and Ullman (AHU) \cite{lca:ahu}, which runs in $O(n + m \alpha(m + n, n))$
time.
The AHU algorithm uses a DSU data
structure; it runs in $O(n + m)$ time on a RAM if this structure is
implemented using the DSU algorithm of Gabow and
Tarjan~\cite{dsu:gt} for the special case in which the set of unions
is known in advance. The first linear-time RAM algorithm was
actually given by Harel and Tarjan~\cite{nca:ht}.  Other linear-time
RAM algorithms were given by Schieber and Vishkin~\cite{lca:sv},
Bender and Farach-Colton~\cite{lca:bf00}, and Alstrup et
al.~\cite{lca:agkr02}.

There are several variants of the NCAs problem of increasing
difficulty.  For each but the last, there is a non-constant-factor
gap between the running time of the fastest RAM and pointer-machine
algorithms.

\begin{description}
\item[Static On-Line:] $T$ is given a priori but $P$ is given on-line:
each NCA query must be answered before the next one is known.
\item[Linking Roots:] $T$ is given dynamically.  Specifically, $T$ is
initially a forest of singleton nodes.  Interspersed with the
on-line NCA queries are on-line $\mlink{v,w}$ operations, each of
which is given a pair of distinct roots $v$ and $w$ in the current
forest and connects them by making $v$ the parent of $w$.
\item[Adding Leaves:] Like linking roots, only $v$ is any node other
than $w$ and $w$ is a singleton.
\item[General Linking:] Like linking roots, only $v$ can be any node
that is not a descendant of $w$.
\item[Linking and Cutting:] Like general linking, but with additional
interspersed $\cut{v}$ operations, each of which is given a non-root
node and makes it a root by disconnecting it from its parent.
\end{description}

Harel and Tarjan \cite{nca:ht} showed that the static on-line
problem (and thus the more general variants) takes $\Omega(\log{\log{n}})$
time on a pointer machine for each query in the worst case.  Alstrup
and Thorup \cite{dnca:at00} gave a matching $O(n +
m\log{\log{n}})$-time pointer-machine algorithm for general linking,
which is also optimal for the static on-line, linking roots, and
adding leaves variants. Earlier, Tsakalidis and van Leeuwen
\cite{nca:tv88} gave such an algorithm for the static on-line
variant, and a modified version of van Leeuwen's even-earlier
algorithm \cite{nca:v76} has the same bound for linking roots.  The
fastest known pointer-machine algorithm for linking and cutting is
the $O(n + m\log{n})$-time algorithm of Sleator and Tarjan
\cite{dynamict:st};
Harel and Tarjan \cite{nca:ht} conjectured that
this is asymptotically optimal,
and the results of P{\v{a}}tra{\c{s}}cu and Demaine \cite{loglb:pd06}
actually imply that lower bound in the cell-probe model.
On a RAM, the fastest known
algorithms take $\Theta(n + m)$ time for the static on-line
\cite{nca:ht,lca:sv} and adding leaves \cite{nca:g} variants, $O(n +
m\alpha(m + n, n))$ time for linking roots \cite{nca:ht} and general
linking \cite{nca:g}, and $O(n + m\log{n})$ time for linking and
cutting \cite{dynamict:st}. All these algorithms use $O(n + m)$
space. For a more thorough survey of previous work see Alstrup et
al. \cite{lca:agkr02}.

\subsection{Verification and Construction of Minimum Spanning
Trees}

\begin{description}
\item[MST Construction:] Given an undirected, connected graph $G = (V,
E)$ whose edges have real-valued weights, find a spanning tree of
minimum total edge weight (an MST) of $G$.
\item[MST Verification:] Given an undirected, connected graph $G = (V, E)$
whose edges have real-valued weights and a spanning tree $T$ of $G$,
determine whether $T$ is an MST of $G$. \end{description}

In both problems, we denote by n and m the numbers of vertices and
edges, respectively.  Since $G$ is connected and $n
\ge 2$, $m \ge n-1$ implies $n = O(m)$.

MST construction has perhaps the longest and richest history of any
network optimization problem; 
Graham and Hell \cite{msthist:gh}
and Chazelle \cite{mst:c00} provide excellent surveys.  
A sequence of faster-and-faster algorithms
culminated in the randomized linear-time algorithm of Karger, Klein,
and Tarjan~\cite{mst:kkt}. This algorithm requires a RAM, but only
for a computation equivalent to MST verification.  It is also
\emph{comparison-based}: the only operations it does on edge weights
are binary comparisons. Previously, Fredman and
Willard~\cite{mstj:fw} developed a linear-time RAM algorithm that is
not comparison-based. Subsequently, Chazelle~\cite{mst:c00}
developed a deterministic, comparison-based $O(m\alpha(m, n))$-time
pointer-machine algorithm, and Pettie and Ramachandran~\cite{mst:pr02}
developed a deterministic, comparison-based pointer-machine
algorithm that runs in minimum time to within a constant factor.
Getting an asymptotically tight bound on the running time of this
algorithm remains an open problem.

Although it remains open whether there is a comparison-based,
deterministic linear-time MST construction algorithm, even for a
RAM, such algorithms do exist for MST verification. Tarjan
\cite{pathcomp:t} gave a comparison-based, deterministic $O(m
\alpha(m, n))$-time pointer machine algorithm for verification.
Koml\'os \cite{mstver:k} showed how to do MST verification in $O(m)$
comparisons, without providing an efficient way to determine which
comparisons to do. Dixon, Rauch, and Tarjan \cite{mst:drt:j}
combined Tarjan's algorithm, Koml\'os's bound, and the tree
partitioning technique of Gabow and Tarjan \cite{dsu:gt} to produce
a comparison-based, deterministic linear-time RAM algorithm.  King
later gave a simplified algorithm \cite{mstver:j:king}.

\subsection{Interval Analysis of Flowgraphs}

A \emph{flowgraph} $G = (V, E, r)$ is a directed graph with a
distinguished \emph{root vertex} $r$ such that every vertex is
reachable from $r$.  A \emph{depth-first spanning tree} $D$ of $G$
is a spanning tree rooted at $r$ defined by some depth-first search
(DFS) of $G$, with the vertices numbered from $1$ to $n$ in preorder
with respect to the DFS (the order in which the search first visits
them).  We identify vertices by their preorder number.  We denote by
$n$ and $m$ the number of vertices and edges of $G$, respectively.

\begin{description}
\item[Interval Analysis:] Given a flowgraph $G$ and a depth-first
spanning tree $D$ of $G$, compute, for each vertex $v$, its
\emph{head} $h(v)$, defined by
\[
\begin{split}
h(v)  =  \max\{ u  : \ & u \text{ is a proper ancestor of }
v \text{ in } D \text{ and there is a path from } v \text{ to } u \text{ in } G \\
& \text{containing only descendants of } u \}, \text{ or null if this set is empty.}
 \\
\end{split}
\]
\end{description}

The heads define a forest called the \emph{interval forest} $H$, in
which the parent of a vertex is its head.  If $v$ is any vertex, the
descendants of $v$ in $H$ induce a strongly connected subgraph of
$G$,
which is called an \emph{interval}; these intervals impose a hierarchy on
the loop structure of $G$.  Interval analysis has been used in
global flow analysis of computer programs \cite{aho:dragon2}, in
testing flowgraph reducibility \cite{reducibility:tarjan}, and in
the construction of two maximally edge-disjoint spanning trees of a
flowgraph \cite{st:t}.  Tarjan~\cite{st:t} gave an $O(m\alpha(m,
n))$-time pointer-machine algorithm for interval analysis using DSU.
The Gabow-Tarjan DSU algorithm \cite{dsu:gt} reduces the running
time of this algorithm to $O(m)$ on a RAM.

\subsection{Finding Dominators}

Let $G = (V, E, r)$ be a flowgraph.  We denote by $n$ and $m$ the
number of vertices and edges of $G$, respectively.  Vertex $v$
\emph{dominates} vertex $w$ if every path from $r$ to $w$ contains
$v$,
and $v$ is the {\em immediate dominator} of $w$ if every
vertex that dominates $w$ also dominates $v$. 
The dominators define a tree rooted at $r$, the \emph{dominator
tree} $T$, such that $v$ dominates $w$ if and only if $v$ is an
ancestor of $w$ in $T$: for any vertex $v \neq r$, the
immediate dominator of $v$ is its parent in $T$.

\begin{description}
\item[Finding Dominators:] Given a flowgraph $G = (V, E, r)$, compute
the immediate dominator of every vertex other than $r$.
\end{description}

Finding dominators in flowgraphs is an elegant problem in graph
theory with fundamental applications in global flow analysis and
program optimization
\cite{ptc:ii:au,flowgraph:c+,flowgraph:fow,domin:lm} and additional
applications in VLSI design \cite{amyeen:01:vlsitest}, theoretical
biology \cite{foodwebs:ab04,foodwebs:abb} and constraint programming
\cite{QVDR:PADL:2006}. Lengauer and Tarjan \cite{domin:lt} gave a
practical $O(m\alpha(m, n))$-time pointer-machine algorithm, capping
a sequence of previous improvements
\cite{ptc:ii:au,domin:lm,domin:pm,domin:tarjan}.
Harel~\cite{domin:harel} claimed a linear-time RAM algorithm, but
Alstrup et al.~\cite{domin:ahlt99} found problems with some of his
arguments and developed a corrected algorithm, which uses powerful
bit-manipulation-based data structures. Buchsbaum et
al.~\cite{domin:bkrw} proposed a simpler algorithm, but Georgiadis
and Tarjan~\cite{dom:gt04} gave a counterexample to their
linear-time analysis and presented a way to repair and modify the
algorithm so that it runs in linear time on a pointer machine;
Buchsbaum et al.~\cite[\em Corrig.]{domin:bkrw} gave a different resolution that
results in a linear-time algorithm for a RAM.

\subsection{Building a Component Tree}

Let $T$ be a tree and let $L$ be a list of the edges of $T$.  The
\emph{Kruskal tree} of $T$ with respect to $L$ is a tree
representing the connected components formed by deleting the edges
of $T$ and adding them back one-at-a-time in the order of their
occurrence in $L$. Specifically, $K$ contains $2n - 1$ nodes.  Its
leaves are the nodes of $T$.  Each internal node is a component
formed by adding an edge $(v,w)$ back to $T$; its children are the
two components that combine to form it.

\begin{description}
\item[Component-Tree Construction:] Given an $n$-node tree $T$ and a list $L$
of its edges, build the corresponding Kruskal tree.
\end{description}

Compressed component trees (formed by adding edges a-group-at-a-time
rather than one-at-a-time) have been used in shortest-path
algorithms \cite{sp:pr02,usp:t99}.  
It is straightforward to build a component tree or a
compressed component tree in $O(n \alpha(n, n))$ time on a pointer
machine using DSU.  The Gabow-Tarjan DSU algorithm \cite{dsu:gt}
improves this algorithm to $O(n)$ time on a RAM,
as described by Thorup \cite{usp:t99}.

\section{Path Compression on Balanced Trees}
\label{sec:pc}

\subsection{Disjoint Set Union Via Path Compression and Balanced
Unions}

The \emph{disjoint set union} (DSU) problem calls for the
maintenance of a dynamic partition of a universe $U$, initially
consisting of singleton sets.  Each set has a unique
\emph{designated element}; the designated element of a singleton set
is its only element.  Two operations are allowed:

\begin{description}
\item[$\unite(v,w)$] Form the union of the sets whose designated
elements are $v$ and $w$, with $v$ being the designated element of
the new set.
\item[$\find(v)$] Return the designated element of the set containing
element $v$.
\end{description}

There are alternative, equivalent formulations of the DSU problem.
In one \cite{setunionptr:tarjan,setunion:tarjan}, 
each set is accessed by a label, rather than by a designated
element.  
In another \cite{setunion:tvl}, sets have labels but can be accessed by
{\em any} element.
In yet another
\cite{setunion:tvl}, each set is accessed by a {\em canonical element},
which in the case of a $\unite(v,w)$ operation can be freely chosen by
the implementation to be either $v$ or $w$.
Our formulation more closely matches our uses.  We denote
by $n$ the total number of elements and by $m$ the total number of
finds.

The standard solution to the DSU problem \cite{setunion:tarjan,setunion:tvl} represents the
sets by rooted trees in a forest.  Each tree represents a set, whose
elements are the nodes of the tree.  Each node has a pointer to its
parent and a bit indicating whether it is a root; 
the root points to the designated element of the set.  To
provide constant-time access to the root from the designated node,
the latter is either the root itself or a child of the root.  With
this representation, to perform $\unite(v,w)$: find the roots of
the trees containing $v$ and $w$, link them together by making one
root the parent of the other, and make $v$ a child of the new root
if it is not that root or a child of that root already.  To perform $\find(v)$:
follow parent pointers until reaching a root, reach the designated
element of the set in at most one more step, and return this
element. A unite operation takes $O(1)$ time.  A find takes time
proportional to the number of nodes on the find path.  A sequence of
intermixed unite and find operations thus takes $O(n + s)$ time,
where $s$ is the total number of nodes on find paths.

One way to reduce $s$ is to use \emph{path compression}: after a
find, make the root the parent of every other node on the find path.
Another way to reduce $s$ is to do \emph{balanced unions}.  There
are two well known balanced-union rules.  In the first,
\emph{union-by-size}, each root stores the number of its
descendants.  To perform $\unite(v,w)$, make the root of the larger
tree the parent of the root of the smaller, making either the 
parent of the other
in case of a tie.  In the second, \emph{union-by-rank}, each root
has a non-negative integer \emph{rank}, initially zero.  To perform
$\unite(v,w)$, make the root of higher rank the parent of the root
of lower rank; in case of a tie, make either root the parent of the other and add
one to the rank of the remaining root.  Both of these union rules produce
\emph{balanced} trees. More specifically, let $F$ be the forest
built by doing all the unite operations and none of the finds.  We
call $F$ the \emph{reference forest}. $F$ is \emph{balanced}, or,
more precisely, \emph{$c$-balanced} if for a constant $c > 1$ the
number of nodes of height $h$ in $F$ is $O(n/c^h)$ for every $h$.
Both union-by-size and union-by-rank produce 2-balanced forests.
Furthermore, since only roots must maintain sizes or ranks,
these fields obviate the need for separate bits
to indicate which nodes are roots.

For any sequence of unions and finds such that the unions build a
balanced forest and the finds use path compression, the total
running time is $O(n + m \alpha(m + n, n))$: 
the analysis of path compression by Tarjan and van
Leeuwen~\cite{setunion:tvl} applies if the reference forest is balanced.
We seek a linear time bound, which we can obtain for sequences of
finds that are suitably restricted. Before obtaining this bound, we
discuss a more general use of path compression and balanced union:
to find minima on paths in dynamic trees.

\subsection{Finding Minima on Paths}
\label{sec:link-eval}

The \emph{dynamic path-minimum problem} calls for the maintenance of
a forest of rooted trees, each initially a one-node tree, whose arcs,
which are directed from parent to child,
have real values.  The trees are subject to three operations:

\begin{description}
\item[$\mlink{v,w, x}$] Nodes $v$ and $w$ are the roots of different
trees in $F$, and $x$ is a real number.  Make $v$ the parent of $w$
by adding arc $(v,w)$ to $F$, with value $x$.
\item[$\findroot{v}$] Return the root of the tree in $F$ containing
the node $v$.
\item[$\meval{v}$] Return the minimum value of an arc on
the path to $v$ from the root of the tree containing it.
\end{description}

We shall denote by $n$ the total number of nodes and by $m$ the
total number of $\ofindroot$ and $\oeval$ operations.  Variants of
this problem include omitting the $\ofindroot$ operation, replacing
minimum by maximum, and requiring the $\oeval$ operation to return
an arc of minimum value rather than just the minimum value.  The two
solutions to be described are easily modified to handle these
variants.  We call a data structure that solves the dynamic
path-minimum problem a \emph{link-eval structure}.

Tarjan\cite{pathcomp:t} considered this problem and developed two
data structures to solve it: a simple one \cite[Sec.~2]{pathcomp:t},
which uses path compression on the forest defined by the
links, and a sophisticated one \cite[Sec.~5]{pathcomp:t}, which
uses path compression on a balanced forest related to the one
defined by the links.  Tarjan's simple link-eval structure uses a
compressed version of $F$, represented by parent pointers, with the
nodes rather than the arcs storing values.  Each root has value
infinity. Perform $\mlink{v,w, x}$ by making $v$ the parent of $w$
and giving $w$ the value $x$. Perform $\findroot{v}$ by following
parent pointers from $v$ to the root of the tree containing it,
compressing this path, and returning the root. Perform $\meval{v}$
by following parent pointers from $v$ to the root of the tree
containing it, compressing this path, and returning the value of
$v$. To compress a path $v_0, v_1, \ldots ,v_k$ with $v_i$ the
parent of $v_{i+1}$ for $0 \le i < k$, repeat the following step for
each $i$ from 2 through $k$: replace the parent of $v_i$ by $v_0$,
and replace the value of $v_i$ by the value of $v_{i-1}$ if the latter
is smaller. Compression preserves the results of $\ofindroot$ and
$\oeval$ operations while making tree paths shorter.

If the final forest $F$ is balanced, then this simple link-eval
structure takes $O(n + m \alpha(m + n, n))$ time to perform a
sequence of operations \cite{pathcomp:t}: the effect of a
compression on the structure of a tree is the same whether the
compression is due to a $\ofindroot$ or an $\oeval$. In
our MST application the final forest is actually balanced. Our
application to finding dominators requires Tarjan's sophisticated
link-eval structure.

\subsection{Delayed Linking with Balancing}
\label{sec:link-by-size}

Tarjan's sophisticated structure delays the effect of some of the
links so that they can be done in a way that makes the resulting
forest balanced.  Since our analysis requires some knowledge of the
inner workings of this structure, we describe it here.  We
streamline the structure slightly, and we add to it the ability to
do $\ofindroot$ operations, which were not supported by the
original. We also describe (in Section \ref{sec:link-by-rank}) a
variant that uses linking-by-rank; the original uses
linking-by-size.

We represent the forest $F$ defined by the link operations by a
\emph{shadow forest} $R$.  Each tree in $F$ corresponds to a tree in
$R$ with the same vertices and the same root.  Each tree $T$ in $R$
is partitioned into one or more subtrees $S_0, S_1, \ldots, S_k$,
such that the root of $S_i$ is the parent of the root of $S_{i+1}$
for $0 \le i < k$, and the root of $S_0$ is the root of $T$. 
We call the roots of the subtrees $S_0,S_1,\ldots,S_k$ (including
the root of $S_0$) {\em subroots}.
We
represent $R$ by a set of parent pointers that are defined for nodes
that are not subroots and, for each subroot, a pointer to
its child that is a subroot, if any. (Each subroot has a
null parent pointer; the deepest subroot has a null child
pointer.) Since parents are needed only for nodes that are not
subroots and child pointers are required only for subroots, 
we can use a single pointer per node to store both kinds of
pointers, if we mark each node to indicate whether it is a subroot.
We shall use $\shp{v}$ to denote the parent of $v$ in its
subtree and $\shc{v}$ to denote the child of $v$ that is a subroot,
if there is one; $\shp{v} = \nul$ if $v$ is a subroot; $\shc{v} =
\nul$ if $v$ is a subroot without a child that is a subroot.

With each node $v$ we store a value $b(v)$.  We manipulate the trees
of $R$ and the node values to preserve two related invariants:
\begin{itemize}
\item[(i)] $\meval{v} = \min \{ b(u) : u \ \mbox{is an ancestor in} \ R \ \mbox{of} \ v \mbox{, and} \ u \ \mbox{is in the
same subtree as} \ v \}$;
\item[(ii)] $b(\shc{v}) \le b(v)$ if $\shc{v}\not=\nul$.
\end{itemize}
To help keep evaluation paths short, we use both path compression
and a variant of union-by-size.  
We denote by $\size{v}$ the number of descendants of $v$ in $R$
and by $\subsize{v}$ the number of descendants of $v$ in the same subtree as $v$.
For convenience, we let $\size{\nul}=0$.
Then $\subsize{v}=\size{v}$ if $v$ is not a subroot,
and $\subsize{v}=\size{v}-\size{\shc{v}}$ if $v$ is a subroot.
We maintain sizes but only for subroots,
which allows us to compute the subsize of a subroot in constant time.

\begin{figure}[t]
\begin{center}
\scalebox{0.8}[0.8]{\input{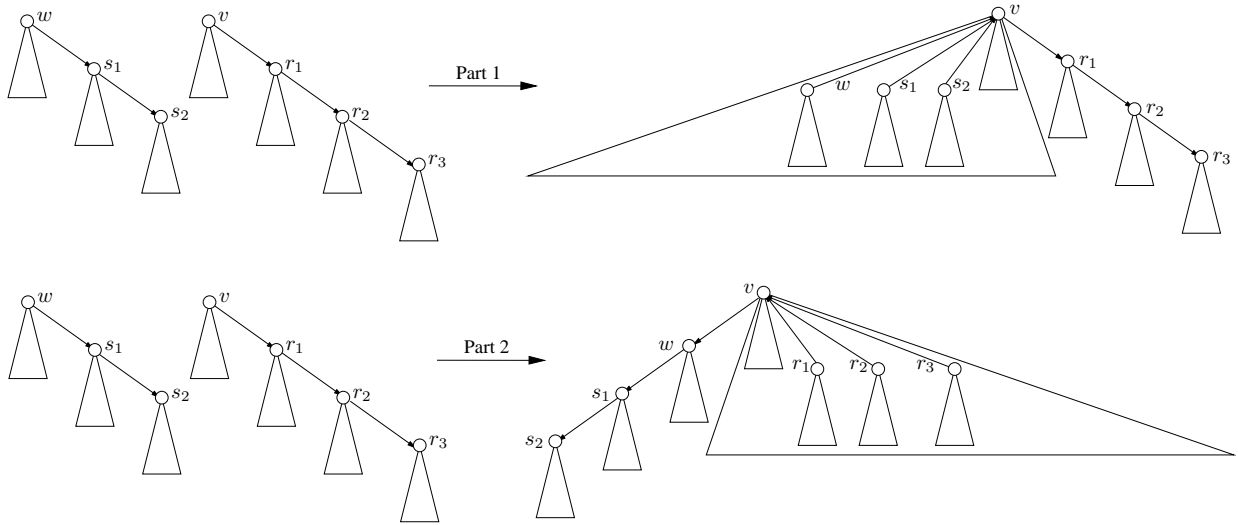}}
\end{center} \caption{Linking by size: Part 1, $\size{v} \ge \size{w}$,  and
Part 2, $\size{v} < \size{w}$. \label{fig:link-by-size}}
\end{figure}

\begin{figure}[t]
\begin{center}
\scalebox{0.8}[0.8]{\input{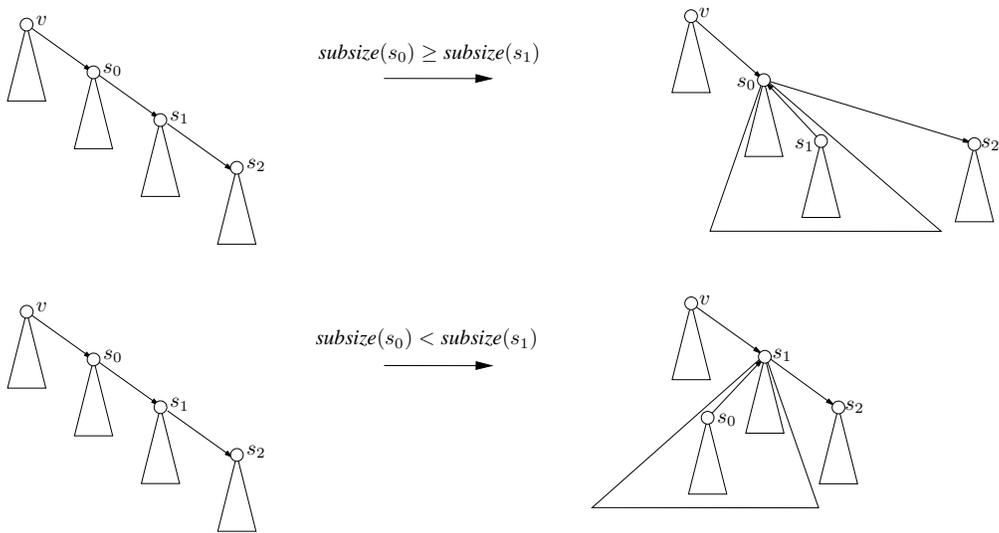}}
\end{center} \caption{Linking by size: Part 3. \label{fig:link-by-size-2}}
\end{figure}

To initialize the structure, make each node $v$ a singleton tree
($\shp{v} = \shc{v} = \nul$), with $b(v) = \infty$ and $\size{v} =
1$. To perform $\meval{v}$, return $b(v)$ if $\shp{v} = \nul$;
otherwise, compress the path to $v$ from the subroot of the subtree
containing it (exactly as in the simple link-eval structure of
Section \ref{sec:link-eval}), and then return $\min\{b(v), b(\shp{v})\}$.
Perform $\mlink{v, w, x}$ as follows. First, set $b(w)$ (previously
infinity) equal to $x$.
Next, if $\size{v}\geq \size{w}$,
perform Part 1 below;
otherwise, perform Part 2 below and, if necessary, Part 3.
(See Figures \ref{fig:link-by-size} and \ref{fig:link-by-size-2}.)
\begin{description}
\item{Part 1:} ($\size{v} \ge \size{w}$.) Combine the subtree rooted
at $v$ with all the subtrees in the tree rooted at $w$, by setting
$\shp{u} = v$ and $b(u) = \min\{b(u), x\}$ for each subroot $u$ of a
subtree in the tree rooted at $w$.  Find such subroots by following
$\mathit{shc}$ pointers from $w$.  
(In Figure \ref{fig:link-by-size}(Part 1), 
the successive values of $u$ are $w,s_1,s_2$.)
This step effects a compression
to $v$ from the deepest subroot descendant of $w$. The updates to
the $b$-values maintain (i) and (ii).
\item{Part 2:} ($\size{v} < \size{w}$.) Combine all the subtrees in
the tree rooted at $v$, by setting $\shp{u} = v$ for each subroot $u
\neq v$ of a subtree in the tree rooted at $v$. 
(In Figure \ref{fig:link-by-size}(Part 2),
the successive values of $u$ are $r_1,r_2,r_3$.)
This step effects a
compression to $v$ from the deepest subroot descendant of $v$. Then
set $\shc{v} = w$. This may cause violations of Invariants (i) and (ii).
\item{Part 3:} 
In order to restore (i) and (ii) after Part 2, repeat the following
step until it no longer applies. 
Let $s_0=\shc{v}$ and $s_1=\shc{s_0}$.
(In the first iteration, $s_0=w$.)
If $s_1\not=\nul$ and $x<b(s_1)$,
compare the subsizes of $s_0$ and $s_1$.
If the former is not smaller,
combine the subtrees
with subroots $s_0$ and $s_1$, making $s_0$ the
new subroot, by simultaneously setting $\shp{s_1} =
s_0$ and $\shc{s_0} = \shc{s_1}$.  If the former
is smaller, combine the subtrees with subroots $s_0$ and
$s_1$, making $s_1$ the new subroot, by
simultaneously setting $\shp{s_0} = s_1$, $\shc{v} =
s_1$, $b(s_1) = x$, and $\size{s_1} =
\size{s_0}$. Once this step no longer
applies, (i) and (ii) are restored.
\end{description}

\noindent
Complete the linking by setting $\size{v}=\size{v}+\size{w}$.
We call this linking method \emph{linking-by-size}.

The method must keep track of which nodes are subroots.
Nodes that are not subroots can be marked as such by, e.g., setting their sizes
to zero,
since sizes are maintained only for subroots.
We have omitted this updating from Parts 1, 2, and 3.

This version of the data structure differs from the original
\cite{pathcomp:t} only in the placement of Part 3 of the link
operation.  In the original, Part 3 is done before Parts 1 and 2 to
restore (i) and (ii), whether or not $\size{v} \ge \size{w}$.
Delaying Part 3 allows it to be avoided entirely if $\size{v} \ge
\size{w}$; in this case Part 1 alone suffices to restore (i) and
(ii).

This structure does not support $\ofindroot$ (because an $\oeval$
operation reaches only a subroot, not a root), but we can easily
extend it to do so.  To each subroot that is not a root, we add a
pointer to its deepest subroot descendant; to each deepest subroot,
we add a pointer to the root of its tree.  Then a root is reachable
from any subroot descendant in at most two steps.  To perform
$\findroot{v}$, compress the path to $v$ from the subroot of its
subtree; then follow at most two pointers to reach a root, and
return this root. Operation $\ofindroot$ has the same asymptotic
complexity as $\oeval$. The running time of a link operation
increases by at most a constant factor because of the extra pointer
updates needed.

In the sophisticated link-eval structure, path compression is 
performed on the
subtrees, not on the trees.  The next lemma implies that these subtrees
are balanced.

\begin{lemma}
Consider a shadow forest built using linking-by-size.  If $u$ is a
tree node such that $\shp{u}$ and $\shp{\shp{u}}$ are both non-null,
then $\subsize{\shp{\shp{u}}} \ge 2 \cdot\subsize{u}$.
\end{lemma}

\begin{proof}
A node $u$ can be assigned a parent $\shp{u}$ in Part 1, 2, or 3 of
a link operation.  If this occurs in Part 3, $\subsize{\shp{u}} \ge
2\cdot\subsize{u}$ after $u$ gets its parent.  Once this happens,
$\subsize{u}$ stays the same and $\subsize{\shp{u}}$ can only
increase. Thus when $\shp{u}$ gets a parent,
$\subsize{\shp{\shp{u}}} \ge \subsize{\shp{u}} \ge 2\cdot\subsize{u}$,
and this inequality persists. 
Regardless of when $u$ gets a parent $\shp{u}$, if $\shp{u}$ gets its
parent in Part 3, then $\subsize{\shp{\shp{u}}} \ge
2\cdot\subsize{\shp{u}} \ge 2\cdot\subsize{u}$ when this happens, and this
inequality persists. Suppose then that both $u$ and $\shp{u}$ get their
parents in Part 1 or 2. When $u$ gets its parent, $\size{\shp{u}}
\ge 2\cdot\subsize{u}$.  Subsequently, $\size{\shp{u}}$ cannot decrease
until $\shp{u}$ gets its parent, at which time
$\subsize{\shp{\shp{u}}} \ge \size{\shp{u}} \ge 2\cdot\subsize{u}$. This inequality
persists.
\end{proof}

\begin{corollary}
\label{cor:link-by-size} The subtrees in any shadow forest built
using linking by size are $\sqrt{2}$-balanced.
\end{corollary}

\subsection{Linking by Rank}
\label{sec:link-by-rank}

An alternative to using linking-by-size in the sophisticated
link-eval structure is to use linking-by-rank.  In place of a size,
every node has a non-negative integer \emph{rank}, initially
zero. The ranks satisfy the invariant

\begin{itemize}
\item[(iii)] $\rank{\shp{v}} > \rank{v}$.
\end{itemize}

We explicitly maintain ranks only for subroots.
If $v$ is a virtual tree root (i.e., in $F$), we denote by $\maxrank{v}$ the
maximum rank of a subroot descendant. 
With each virtual tree root $v$, we
store $\maxrank{v}$ (in addition to $\rank{v}$).

\begin{figure}[t]
\begin{center}
\scalebox{0.775}[0.775]{\input{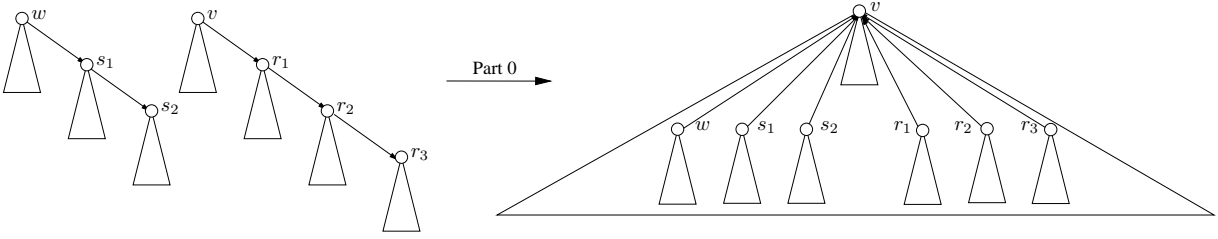}}
\end{center} \caption{Linking by rank: Part 0, $\maxrank{v} =
\maxrank{w}$. \label{fig:link-by-rank}}
\end{figure}

Perform $\mlink{v, w, x}$ as follows.  First, set $b(w) = x$.  Then
compare $\maxrank{v}$ to $\maxrank{w}$.  We split the rest of the
operation into the following parts.
\begin{description}
\item{Part 0:} If $\maxrank{v} =
\maxrank{w}$, set $\rank{v} = \maxrank{v} + 1$,
$\maxrank{v}=\maxrank{v}+1$, and combine all the
subtrees in the trees rooted at $v$ and $w$ into a single subtree
rooted at $v$, by setting $\shp{u} = v$ for each subroot $u \neq v$,
setting $\shc{v}=\nul$,
and setting $b(u) = \min \{ b(u), b(w) \}$ if $u$ was a descendant
of $w$.
(See Figure \ref{fig:link-by-rank}.)
\item{Part 1:} If $\maxrank{v}
> \maxrank{w}$, set $\rank{v} = \max \{ \rank{v}, \maxrank{w} + 1
\}$, and combine the subtree rooted at $v$ with all the subtrees in
the tree rooted at $w$, by setting $\shp{u} = v$ and $b(u) =
\min\{b(u), b(w)\}$ for each subroot descendant $u$ of $w$.
\item{Part 2:} If $\maxrank{v} < \maxrank{w}$,
combine all the subtrees in the tree rooted at $v$ into a single
subtree, unless $\shc{v} = \nul$, by setting $\rank{v} = \maxrank{v}
+ 1$,
\maxrank{v}=\maxrank{w}, and, for each subroot $u \neq v$, $\shp{u} = v$. Then set
$\shc{v} = w$.
This may cause violations of Invariants (i) and (ii).
\item{Part 3:} To restore (i) and (ii) after Part 2,
repeat the following step until it no longer applies.
Let $s_0=\shc{v}$ and $s_1=\shc{s_0}$.
If $s_1\not=\nul$ and $x<b(s_1)$, compare
$\rank{s_0}$ to
$\rank{s_1}$,
and:
if $\rank{s_0} =
\rank{s_1}$, simultaneously set $\shp{s_1} =
s_0$, $\shc{s_0} = \shc{s_1}$, $\rank{s_0}
= \rank{s_0} + 1$,
and $\maxrank{v}=\max\{\maxrank{v},\rank{s_0}+1\}$; 
if $\rank{s_0} > \rank{s_1}$,
simultaneously set $\shp{s_1} = s_0$ and
$\shc{s_0} = \shc{s_1}$; if $\rank{s_0} <
\rank{s_1}$, simultaneously set $\shp{s_0} =
s_1$, $\shc{v} = s_1$, and $b(s_1) = x$.
\end{description}

Parts 1, 2, and 3 of linking-by-rank correspond to Parts 1, 2, and 3
of linking-by-size; Part 0 handles the case of equal
$\mathit{maxranks}$, in which all subtrees of both trees are
combined. (We could add a corresponding Part 0 to linking-by-size,
but this is unnecessary.) As does linking-by-size, linking-by-rank
produces balanced forests, as we now show.  For a node $u$, let
$\subsize{u}$ be the number of descendants of $u$ in its subtree.

\begin{lemma}
In any shadow forest built using linking-by-rank, any node $u$ has
$\subsize{u} \ge 2^{(\rank{u} - 1)/2}$.
\end{lemma}

\begin{proof}
To obtain this result we actually need to prove something stronger.
Suppose we perform a sequence of link-by-rank operations. We track
the states of nodes, their ranks, and their subsizes as
the links take place.  Each node is in one of two states:
\emph{normal} or \emph{special}.  The following invariants will
hold:
\begin{itemize}
\item[(a)] a {\em normal node} $u$ has $\subsize{u} \ge 2^{\rank{u}/2}$;
\item[(b)] a {\em special node} $u$ has $\subsize{u} \ge 2^{(\rank{u} - 1)/2}$;
\item[(c)] a {\em special root} $u$ has a normal subroot descendant of
rank at least $\rank{u}$.
\end{itemize}
Initially all nodes are normal; since all initial ranks are zero,
(a), (b), and (c) hold initially.  We need to determine the effect
of each part of an operation $\mlink{v, w, x}$.

If $\maxrank{v} = \maxrank{w}$, we make $v$ normal after the link;
all other nodes retain their states. This preserves (a), (b), and
(c); the only question is whether $v$ satisfies (a), since it gains
one in rank and can change from special to normal.  Before the link,
both the tree rooted at $v$ and the tree rooted at $w$ have a
subroot of rank $\maxrank{v}$.
Since each of these nodes has subsize at least $2^{(\maxrank{v}-1)/2}$
before the link by (a) and (b),
after the link $\subsize{v}\geq 2\cdot2^{(\rank{v}-2)/2}=2^{\rank{v}/2}$.
Hence (a) holds for $v$ after the link.

If $\maxrank{v} > \maxrank{w}$ and $\rank{v}$ does not change as a
result of the link, all nodes retain their states.  The link
preserves (a), (b), and (c), because no node increases in rank.
If $\rank{v}$ does change because of
the link (becoming one greater than the old value of $\maxrank{w}$),
we make $v$ special. Node $v$ now satisfies (b), because before the
link $w$ had a normal subroot descendant $u$ of rank $\maxrank{w}$,
and $\subsize{u} \ge 2^{\maxrank{w}/2}$ by (a); hence, after the
link, $\subsize{v} \ge 2^{(\rank{v} - 1)/2}$.  Node $v$ satisfies
(c), because before the link it had a normal subroot descendant $z$
of rank $\maxrank{v} \ge \maxrank{w} + 1$, which it retains after
the link.

The last case is $\maxrank{v} < \maxrank{w}$.  In this case we look
at the effects of Part 2 and Part 3 separately.  If Part 2 does
anything, we make $v$ special.  Node $v$ satisfies (b), because
before the link it had a normal subroot descendant of rank
$\maxrank{v}$, which satisfied (a); hence, after the link,
$\subsize{v} \ge 2^{(\rank{v} - 1)/2}$.  Node $v$ satisfies (c)
after the link, because before the link $w$ had a normal subroot
descendant of rank $\maxrank{w} \ge \maxrank{v} + 1$ by (a), which
becomes a normal subroot descendant of $v$.  

Finally, we must
account for the effect of Part 3.  Each combination of subtrees done
by Part 3 preserves (a), (b), and (c), except possibly for those
that combine two subtrees with subroots, say $y$ and $z$, of equal
rank. In this case the rank of the surviving subroot  increases by
one; and if the ranks of $y$ and $z$ previously equaled
$\maxrank{v}$, $\maxrank{v}$ increases by one.  To preserve the
invariants in this case, we make the surviving root, say $y$,
normal. Now $y$ satisfies (a), because before the subtrees rooted at
$y$ and $z$ were combined, both $y$ and $z$ have subsize at least
$2^{(\rank{y} - 1)/2}$; after the subtrees are combined,
$\subsize{y} \ge 2 \cdot 2^{(\rank{y} -2)/2} = 2^{\rank{y}/2}$.
Because $y$ satisfies (a), $v$ satisfies (c).

Thus linking preserves the invariants.  By induction, they remain
true throughout any sequence of links.  The lemma follows from (a)
and (b).
\end{proof}

\begin{corollary}
\label{cor:link-by-rank} The subtrees in any shadow forest built
using linking-by-rank are $\sqrt{2}$-balanced.
\end{corollary}

\begin{theorem}
A sequence of operations performed using the sophisticated link-eval
structure with either linking-by-size or linking-by-rank takes
$O(n)$ time for the links and $O(n + m\alpha(m + n, n))$ time for
the $\ofindroot$ and $\oeval$ operations.
\end{theorem}

\begin{proof}
The time for a link is $O(k + 1)$, where $k$ is the decrease
in the number of subtrees caused by the link.  Thus the total time
for all the links is $O(n)$.  The total length of compressed paths,
and hence the total time for $\ofindroot$ and $\oeval$ operations,
is $O(n + m\alpha(m + n, n))$ by the Tarjan-van Leeuwen analysis of
path compression \cite{setunion:tvl}, applying Corollary
\ref{cor:link-by-size} (for linking-by-size) or Corollary
\ref{cor:link-by-rank} (for linking-by-rank).
\end{proof}

\subsection{Refined Analysis of Path Compression}

In order to use path compression on balanced trees as a tool for
building linear-time algorithms, we need to show that the total time
becomes linear if the compressions are suitably restricted.  In
order to capture both DSU and link-eval applications, we abstract
the situation as follows.  An intermixed sequence of the following
two kinds of operations is performed on a rooted forest, initially
consisting of $n$ single-node trees:

\begin{description}
\item[$\assign{u, v}$] Given two distinct roots $u$ and $v$, make $u$ the
parent of $v$.
\item[$\compress{u}$] Compress the path to $u$ from the root of the tree
containing it, by making the root the parent of every other node on
the path.
\end{description}

\begin{lemma}
\label{lemma:pc}
Suppose $\ell$ nodes are {\em marked} and the remaining
$n - \ell$ {\em unmarked}. Suppose the assignments build a balanced
forest, and that each node has its parent change at most $k$ times
before it is in a tree containing a marked node.  
If there are $m$ compress operations, then the total
number of nodes on compression paths is $O(kn + m \alpha(m + \ell,
\ell))$.
\end{lemma}

\begin{proof}
Let $F$ be the balanced forest built by the entire sequence of
assignments, ignoring the compressions; let $c > 1$ be such that $F$
is $c$-balanced; and let $h(v)$ be the height of a node $v$ in $F$.
Let
\begin{eqnarray*}
a = \big \lceil  \log_{c}{(n/\ell)} + \log_{c}{(1/(c - 1))}  + 1
\big \rceil.
\end{eqnarray*}
Classify each node $v$ into one of three types:
\emph{low}, if $v$ has no marked descendant in $F$; \emph{middle},
if $v$ has a marked descendant in $F$ and $h(v) < a$; and
\emph{high} otherwise.

A compression path from a tree root to one of its descendants consists of zero or
more high nodes followed by zero or more middle nodes followed by
zero or more low nodes.   Every node on the path except the first
two (totaling at most $2m$ over all compressions) has its parent
change to one of greater height as a result of the
compression.

Consider a compression path containing only low nodes.  Since the
root is low, the tree in which the compression takes place contains
no marked nodes.  All but two nodes on the path change parent but
remain in a tree with no marked nodes.  The number of times this can
happen to a particular node is at most $k$ by the hypothesis of the
lemma, totaling at most $kn$ over all compressions.

Consider a compression path containing at least one middle or high
node. Every low node on the path except one has its parent change
from low to middle or high as a result of the compression.  Thus the
total number of low nodes on such paths is at most $n + m$.  Every
middle node on the path whose parent changes obtains a parent
of greater height.  This can happen to a middle node at most $a$ times
before its parent is high.  At most one middle node on a compression
path has a high parent, totaling at most $m$ over all compression
paths. Each middle node has a marked node as a descendant; each
marked node has at most $a + 1$ middle nodes as ancestors (at most
one per height less than $a$).  The total number of middle nodes is
thus at most $\ell(a + 1)$.  Combining estimates, we find that the
total number of middle nodes on compression paths is at most $\ell
\cdot a \cdot (a + 1) + m$. Since $\ell \le n$ and $a$ is
$O(\log{(n/\ell)})$, the first term is $O(n)$, implying that the
total number of middle nodes on compression paths is $O(n) + m$.

Finally, we need to count the number of high nodes on compression
paths. Since $F$ is $c$-balanced, the total number of high nodes is
at most
$$
\sum_{i \ge a}{\frac{n}{c^i}} \le \frac{n}{c^a} \cdot \frac{c}{c -
1} = \frac{n}{c^{a - 1}(c - 1)} \le \ell.
$$
Let the \emph{rank} of a node $v$ be $h(v) - a$.  Then every high
node has non-negative rank, and the number of high nodes of rank $i
\ge 0$ is at most $\ell/c^i$.  The analysis of Tarjan and
van-Leeuwen~\cite[Lem.~6]{setunion:tvl} applied to the high
nodes bounds the number of high nodes on compression paths by
$O(\ell + m\alpha(m + \ell, \ell))$.  Combining all our estimates
gives the lemma.
\end{proof}

Lemma \ref{lemma:pc} gives a bound of $O(n+m)$ if, for example, $\ell =
O(n/\log{\log{n}})$, by the properties of the inverse-Ackermann
function \cite{setunion:tarjan}.  In our applications $\ell =
n/\log^{1/3}{n}$, which is sufficiently small to give an $O(n+m)$
bound.

We conclude this section by reviewing some previous results on
disjoint set union and refined analysis of the DSU structure.  The
linear-time RAM DSU algorithm of Gabow and Tarjan~\cite{dsu:gt}
assumes a priori knowledge of the unordered set of unions.  An
earlier version of our work \cite{ptrs:bkrw} contained a result much
weaker than Lemma \ref{lemma:pc}, restricted to disjoint set union,
which required changing the implementation of unite based on the
marked nodes. Alstrup et al.~\cite{domin:ahlt99} also proved a
weaker version of Lemma \ref{lemma:pc} in which the $m\alpha(m+\ell,
\ell)$ term is replaced by $\ell \log{\ell} + m$, which sufficed for
their purpose. They derived this result for a hybrid algorithm that
handles long paths of unary nodes outside the standard DSU
structure. Dillencourt, Samet, and Tamminen~\cite{lindsu:dst} gave a
linear-time result assuming the \emph{stable tree property}:
essentially, once a find is performed on any element in a set $X$,
all subsequent finds on elements currently in $X$ must be performed
before $X$ can be united with another set.  Fiorio and
Gustedt~\cite{ipdsu:fg96} exploit the specific order of unions in an
image-processing application.  Gustedt~\cite{dsu:g98} generalizes
the previous two works to consider structures imposed on sets of
allowable unions by various classes of graphs.  This work is
orthogonal to that of Gabow and Tarjan~\cite{dsu:gt}. Other improved
bounds for path compression \cite{mindeq:bst2,postdsu:j:ln,dsu:l}
restrict the order in which finds are performed, in ways different
from our restriction.

\section{Topological Graph Computations}
\label{sec:tgc}

Consider some computation that takes as input a graph $G$ whose
vertices and edges (or arcs) have $O(1)$-bit labels, and produces
some output information (possibly none) associated with the graph
itself and with each vertex and edge (or arc).  We call such a
computation a \emph{topological graph computation}, because it is
based only on the graph structure and the $O(1)$-bit labels, in
contrast, for example, to a problem in which graph vertices and
edges (or arcs) have associated real values.  In general the output
of a topological graph computation can be arbitrarily complex, even
exponential in size, and can contain pointers to elements of the
input graph.  Our MST verification algorithm will exploit this
flexibility; in all our other applications, the size of the output
is linear in the size of the input.

Suppose we need to perform a topological graph computation on not
just one input graph but on an entire collection of graphs.  If the
input instances are small and there are many of them, then many of
them will be isomorphic.  By doing the computation once for each
non-isomorphic instance (a \emph{canonical instance}) and copying
these solutions to the duplicate instances, we can
amortize away the cost of actually doing the computations on the
canonical instances; most of the time is spent identifying the
isomorphic instances and transferring the solutions from the
canonical instances to the duplicate ones.  The total time spent is
then linear in the total size of all the instances.

Gabow and Tarjan\cite{dsu:gt} used this idea to solve a special case
of disjoint set union in which the unordered set of unions is given
in advance; Dixon et al.\cite{mst:drt} applied the technique to MST
verification and other problems.  These applications use table
look-up and require a RAM. Here we describe how to accomplish the
same thing on a pointer machine.   Our approach is as follows.
Encode each instance as a list of pointers. Use a pointer-based
radix sort to sort these lists. Identify the first instance in each
group of identically-encoded instances as the canonical instance. Solve the
problem for each canonical instance.  Map the solutions back to the
duplicate instances.  The details follow.

Let $\mathcal{G}$ be the set of input instances, each of which
contains at most $g$ vertices.  Let $N$ be the total number of
vertices and edges (or arcs) in all the instances.  Let $k$ be the
maximum number of bits associated with each vertex and edge of an
instance. Construct a singly linked master list whose nodes, in
order, represent the integers from zero through $\max\{g, 2^k + 1\}$
and are so numbered.  For each instance $G$, perform a depth-first
search, numbering the vertices in preorder and adding to each vertex
a pointer into the master list corresponding to its preorder number;
the preorder numbering allows us to maintain a global pointer
into the master list to facilitate this assignment of pointers
to vertices.
Represent the label of each vertex and edge by a pointer into the
master list, using a pointer to the zero node to encode the lack of
a label.  Construct a list $L$ of triples corresponding to the vertices
of $G$, one triple per vertex, consisting of a pointer to the
vertex, and its number and label, both represented as pointers into
the master list.  Construct a list $Q$ of quadruples corresponding
to the edges (or arcs) of the graph, one quadruple per edge (or
arc), consisting of a pointer to the edge (or arc), and the numbers
of its endpoints and its label, represented as pointers into the master
list. (For an undirected graph, order the numbers of the edge
endpoints in increasing order.) Encode the instance by a list whose
first entry is a pair consisting of a pointer to the instance and
the number of its vertices, represented as a pointer into the master
list, catenated with lists $L$ and $Q$. 

Constructing
encodings for all the instances takes $O(N)$ time.  
Recall that the elements of the encodings are pointers
to the master list.
Attach a bucket to each element of the master list.
Use a radix sort for variable length lists \cite{algs:ahu},
following the encoding pointers to reach the buckets,
to arrange the encodings into groups that are identical except for the
first components of each list element (pair, triple, or quadruple):
instances whose encodings are in the same group are isomorphic. This
also takes $O(N)$ time.

Now 
perform the topological graph
computation on any one instance of each group (the canonical instance
for that group).  
Finally, for
each duplicate instance, traverse its encoding and the encoding of
the corresponding canonical instance concurrently, transferring the
solution from the canonical instance to the duplicate instance.  The
exact form this transfer takes depends upon the form of the output
to the topological graph computation.  One way to do the transfer is
to traverse the encodings of the canonical instance and the
duplicate instance in parallel, constructing pointers between
corresponding vertices and edges (or arcs) of the two instances.
Then visit each vertex and edge (or arc) of the canonical instance,
copying the output to the duplicate instance but replacing each
pointer to a vertex or edge (or arc) by a pointer to the
corresponding vertex or edge (or arc) in the duplicate instance.  If
the output has size linear in the input, this takes $O(N)$ time.
Summarizing, we have the following theorem.

\begin{theorem}
\label{thm:tgc} If the output of a topological graph computation has
size linear in the input size, the computation can be done on a
collection of instances of total size $N$ in $O(N)$ time on a
pointer machine, plus the time to do the computation on one instance
of each group of isomorphic instances.
\end{theorem}

This method extends to allow the vertices and edges (or arcs) of the
instances to be labeled with integers in the range $[1, g]$, if
these labels are represented by pointers to the nodes of a
precomputed master list.  We shall need this extension in our
applications to finding dominators and computing component trees
(Sections \ref{sec:dom} and \ref{sec:ct}, respectively). In another
of our applications, MST verification, the output of the topological
graph computation has exponential size: it is a comparison tree,
whose nodes indicate comparisons between the weights of two edges.
In this case, we do not construct a new copy of the comparison tree
for each duplicate instance. Instead, when we are ready to run the
comparison tree for a duplicate instance, we construct pointers
from the edges of the canonical instance to the corresponding edges
of the duplicate instance and run the comparison tree constructed
for the canonical instance, but comparing weights of the
corresponding edges in the duplicate instance.  The total time is
$O(N)$ plus the time to build the comparison trees for the canonical
instances plus the time to run the comparison trees for all the
instances.

It remains to bound the time required to do the topological graph
computation on the canonical instances.  The number of canonical
instances is $g^{O(g^2)}$.  In all but one of our applications, the
time to do a topological graph computation on an instance of size
$g$ or smaller is $O(g^2)$; for MST verification, it is
$g^{O(g^2)}$. Thus the following theorem suffices for us:

\begin{theorem}
\label{thm:tgc2} If a topological graph computation takes
$g^{O(g^2)}$ time on a graph with $g$ or fewer vertices, and if $g =
\log^{1/3}{N}$, then the total time on a pointer machine to do the
topological graph computation on a collection of graphs of total
size $N$, each having at most $g$ vertices, is $O(N)$.
\end{theorem}
\begin{proof}
Immediate from Theorem \ref{thm:tgc}, since the total time to do the
topological graph computation on the canonical instances is
$g^{O(g^2)}g^{O(g^2)} = g^{O(g^2)} = O(N)$.
\end{proof}

The ability to recover the answers from the topological graph computations
on the instances in $\GG$
is subtle yet critical.
Alstrup, Secher, and Spork \cite{deccon:ass}
show how to compute connectivity queries on a tree $T$
undergoing edge deletions in linear time.
They partition $T$ into bottom-level microtrees
(discussed in the next section)
and compute, for each vertex $v$ in a microtree,
a bit-string
that encodes the vertices on the path from $v$ to the root
of its microtree.
They show how to answer connectivity queries
using a constant number of bitwise operations on these bit-strings
and applying the Even and Shiloach 
decremental connectivity algorithm \cite{onlineconn:es}
to the upper part of $T$.

The Alstrup, Secher, and Spork algorithm \cite{deccon:ass} 
runs on a pointer machine:
since the connectivity queries return yes/no answers,
they need not index tables to recover the answers.
In contrast,
while their method can be extended to solve the off-line
NCAs problem in linear time on a RAM,
and even to simplify 
the Gabow-Tarjan linear-time DSU result \cite{dsu:gt},
both of these extensions require indexing tables to map
the results of the bitwise operations
back to vertices in $T$.

The idea of using pointers to buckets in lieu of indexing an array was
described in general by Cai and Paige \cite{disc:cp95} in the context of multi-sequence
discrimination.  Their technique leaves the efficient identification of
buckets with specific elements as an application-dependent problem. 
They solve this problem for several applications, including
discriminating trees and DAGs, but their solutions exploit
structures specific to their applications
and do not extend to general graphs.

\section{Nearest Common Ancestors}
\label{sec:nca}

We now have the tools to solve our first application, the
off-line nearest common ancestors (NCAs) problem: given a rooted
$n$-node tree $T$ and a set $P$ of $m$ queries, each of which is a pair
$\pair{v,w}$ of nodes in $T$, compute $\nca{v, w}$ for each query
$\pair{v,w}$. Aho, Hopcroft, and Ullman's algorithm\cite{lca:ahu}
for this problem, as presented by Tarjan\cite{pathcomp:t}, solves it
using DSU. The algorithm traverses $T$ bottom-up,
building a shadow copy as a DSU forest.
It maintains, for each
subtree built so far, the set of its nodes, with the root of the
subtree as the designated element. Initially, each node is in a
singleton set. Each node $v$ also has a set $P(v)$ of queries
$\pair{v, w}$; each query is in two such lists, one for $v$ and one
for $w$.  The algorithm is as follows.

Visit the nodes of $T$ in a postorder \cite{dfs:t}.  (Any postorder
will do.) When visiting a node $v$, for every pair $\pair{v, w}$ in
$P(v)$ such that $w$ has already been visited, return $\find(w)$ as
the answer to the query \nca{v,w}.
Finish the visit to $v$ by performing $\unite(p(v),v)$ if $v$
is not the root of $T$, where $p(v)$ is the parent of $v$ in $T$.

The correctness of this algorithm follows from basic properties of
postorder.  The DSU operations dominate the running time, which is
$O(n + m\alpha(m+n, n))$ if the standard DSU structure presented in
Section \ref{sec:pc} is used.  In this algorithm, the unordered set
of unions is known in advance, since it is given by the input tree
$T$. Thus the use of the Gabow-Tarjan\cite{dsu:gt} linear-time RAM
DSU algorithm results in a linear-time RAM algorithm for NCAs.
Knowing the set of unions in advance, however, is not sufficient to
solve the DSU problem in linear time on a pointer
machine\cite{dsu:l}. We exploit a different property of the unions:
they occur in a bottom-up order.

\begin{figure}[t]
\begin{center}
\scalebox{0.825}[0.825]{\input{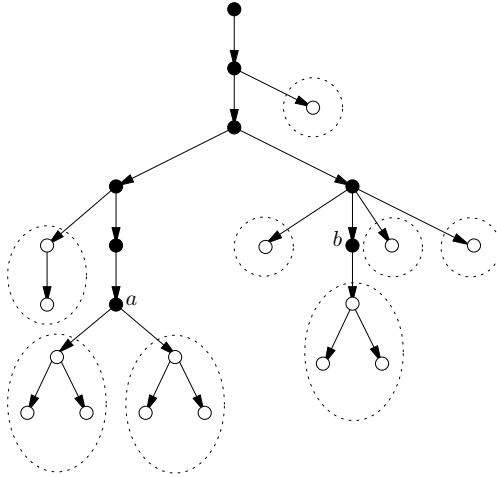}}
\end{center} \caption{Partitioning of a tree $T$
with $g=3$; fringe nodes are open, and
core nodes are filled; bottom-level microtrees are encircled. 
Nodes $a$ and $b$ are the leaves of the core.
\label{fig:dfs}}
\end{figure}

We partition $T$ into a set of small bottom-level trees, called
\emph{microtrees}, and $T'$, the rest of $T$.  For any node $v$,
let $T(v)$ be the subtree of $T$ induced by the descendants of $v$,
and let $|T(v)|$ be the number of nodes in $T(v)$.  Let $g \ge 1$ be
a fixed parameter to be chosen later.  We define $T(v)$ to be a
{\em microtree} if $|T(v)| \le g$ but $|T(p(v))| > g$.  For a node $x$ in
$T(v)$, $\micro{x} = T(v)$ is the \emph{microtree of} $x$ and
$\mroot{\micro{x}}$ is the \emph{root of its microtree}. Let $T'$ be
the subtree of $T$ induced by the vertices in $T$ that are not in
microtrees.  Each leaf in $T'$ has at least $g$ descendants in $T$,
and the descendants of two different leaves of $T'$ form disjoint
sets, so $T'$ has at most $n/g$ leaves.  We call the microtrees the
\emph{fringe} of $T$ and $T'$ the \emph{core} of $T$.  
See Figure \ref{fig:dfs}.
It is
straightforward to partition $T$ into its microtrees and core in
linear time by visiting the nodes in postorder and computing their
numbers of descendants.

We call a query $\pair{v, w}$ \emph{small} if $v$ and $w$ are in the
same microtree and \emph{big} otherwise.  We can partition the
queries into big and small and assign each small query to the
microtree containing it in linear time. We answer all the big
queries by using the AHU algorithm.  We answer all the small queries
by doing a topological graph computation on the set of graphs
defined by each microtree and its associated queries.  By choosing
$g$ appropriately, we get a linear time bound for both parts of the
computation.

Specifically, choose $g = \log^{1/3}{n}$.  Answer all the big
queries by running the AHU algorithm, restricted to the big queries.
To bound the running time, apply Lemma \ref{lemma:pc} to the tree
built by the parent assignments done by the unite operations. Mark
every leaf of $T'$. Each find occurs in a set containing at least
one marked node.
Therefore, setting $k = 1$, to count the initial
parent assignment for each node, satisfies the hypothesis of the
lemma. Since the number of marked nodes is at most $n/g =
n/\log^{1/3}{n}$, the lemma implies an $O(n + m)$ bound on the time
to answer all the big queries.

Answer all the small queries by constructing, for each microtree, 
a
graph containing the microtree arcs
and, for each query with both nodes in the microtree,
an arc denoted as a query arc by a bit.
  Then do a topological graph computation on these graphs
to answer the small queries, using the method of Section
\ref{sec:tgc}. With $g = \log^{1/3}{n}$, this takes $O(n + m)$ time.
Thus we obtain:

\begin{theorem}
\label{thm:nca} The off-line NCAs problem can be solved in $O(n +
m)$ time on a pointer machine.
\end{theorem}

\section{Minimum Spanning Trees}
\label{sec:mst}

\subsection{Verification}

Our next applications, minimum spanning tree (MST) verification and
construction, combine topological graph processing with use of the
simple link-eval structure of Section \ref{sec:link-eval}.  Let $T$
be a spanning tree of a connected, undirected graph $G$ whose edges
have real-valued weights. For any edge $\pair{v, w}$, let $c(v,w)$
be the weight of $\pair{v, w}$. We denote the set of non-tree edges
by $P$. For any pair $(v, w)$ of vertices, we denote by $T(v, w)$
the unique path from $v$ to $w$ in $T$.  The tree $T$ is minimum if
and only if, for every edge $\pair{v, w}$ in $P$, $c(v, w) \ge c(x,
y)$ for every edge $\pair{x,y}$ on $T(v, w)$. Thus to verify that $T$ is minimum
it suffices to compute $\max\{ c(x,y) : \pair{x, y} \ \mbox{on} \
T(v,w)\}$ for every edge $\pair{v,w}$ in $P$. 
We assume henceforth that $T$ is rooted at a fixed but arbitrary
vertex and that each vertex $v$ has a set $P(v)$ of the pairs
$\pair{v, w}$ in $P$.

Tarjan's $O(m\alpha(m, n))$-time MST verification
algorithm\cite{pathcomp:t} is like the AHU NCAs algorithm, except
that it uses a link-eval structure (with $\max$ instead of $\min$)
in place of a DSU structure to compute the needed path maxima.  The
algorithm builds the link-eval forest
during a bottom-up traversal of $T$. As part of the process of
computing path maxima, the algorithm computes $u = \nca{v, w}$ for
each pair $\pair{v, w}$ in $P$ and stores $\pair{v, w}$ in a set
$Q(u)$. Initially each node of $T$ is in a single-node tree of the
link-eval structure, and $Q(u)$ is empty for each node $u$.  
The algorithm follows.

Visit the nodes of $T$ in a postorder.  (Any postorder will do.)  When
visiting a vertex $v$, for every pair $\pair{v, w}$ in $P(v)$ such
that $w$ has already been visited, add $\pair{v, w}$ to
$Q(\findroot{w})$. For every pair $\pair{x, y}$ in $Q(v)$, return
$\max\{\meval{x}, \meval{y}\}$ as the answer to the query $\pair{x,
y}$. Finish the visit to $v$ by performing $\mlink{p(v), v, c(p(v),
v)}$ unless $v$ is the root of $T$.

When the algorithm answers a query $\pair{x, y}$ while visiting a
vertex $v$, $v = \nca{x,y}$, and $\meval{x}$ and $\meval{y}$ are the
maximum costs of the arcs on $T(v, x)$ and $T(v, y)$, respectively.
In Tarjan's original presentation, the NCA calculations are separate
from the path evaluations, but combining them gives a more coherent
algorithm. Ignoring the arc costs and $\oeval$ operations, the
link-eval structure functions exactly like the DSU structure in the
AHU NCAs algorithm.

If the sophisticated link-eval structure of Section
\ref{sec:link-by-size} or Section \ref{sec:link-by-rank} is used,
this algorithm runs in $O(m\alpha(m, n))$ time. Unfortunately, these
structures delay the effect of the links, so parent
assignments do not necessarily occur in a bottom-up order, and we
cannot immediately apply the approach of Section \ref{sec:nca} to
reduce the running time to linear.  This problem was pointed out by
Georgiadis and Tarjan~\cite{dom:gt04}.  Instead, we use a result of
King~\cite{mstver:j:king} to transform the original tree into an
$O(n)$-node balanced tree on which to compute path maxima.  Then we
can use the simple link-eval structure of Section
\ref{sec:link-eval} in combination with the approach of Section
\ref{sec:nca} to obtain a linear-time algorithm.

\subsection{The \boruvka~Tree}

A \emph{\boruvka~step} \cite{mst:bor} applied to a weighted,
undirected graph $G$ is as follows: select a least-weight edge
incident to each vertex, and contract to a single vertex each
connected component formed by the selected edges.  Repeating this
step until only a single vertex remains produces an MST defined by
the original edges corresponding to the edges selected in all the
steps, if all edge weights are distinct, which we can assume without
loss of generality.

This algorithm can be enhanced to produce the \emph{\boruvka~tree}
$B$, whose nodes are the connected components that exist during the
\boruvka~steps, with each node having as children those components
from which it is formed during a \boruvka~step.  If component $C$
is the parent of component $D$, the weight of arc $(C, D)$ is the
weight of the edge selected for the vertex corresponding to $D$ by
the \boruvka~step in which $D$ is contracted into $C$.  The leaves
of $B$ are the vertices of $G$, each of which is originally a
single-vertex component.  Each \boruvka~step reduces the number of
vertices by at least a factor of two; hence, $B$ is 2-balanced. Also,
$B$ contains at most $2n - 1$ nodes.  In general the enhanced
\boruvka~algorithm runs in $O(m\log{n})$ time on a pointer machine.
On a tree, however, it runs in $O(n)$ time, because each contracted
graph is a tree, and a tree has $O(n)$ edges. We apply the enhanced
\boruvka~algorithm to the tree $T$ that is to be verified, thereby
constructing the \boruvka~tree $B$ of $T$.  In addition to being
balanced, $B$ has the following key property\cite{mstver:j:king}:
for any pair of vertices $\pair{v,w}$, $\max\{c(x,y) : (x, y) \
\mbox{on} \ T(v, w)\} = \max\{c(x, y) : (x, y) \ \mbox{on} \
B(v,w)\}$. Thus we can compute path maxima on $B$ instead of on $T$
without affecting the answers to the queries.

\subsection{Comparison Trees for Computing Path Maxima}
\label{sec:mst:path_max} Now we can apply the approach of Section
\ref{sec:nca}. Let $g = \log^{1/3}{n}$. Partition $B$ into
microtrees and a core $B'$ as in Section \ref{sec:nca}. Partition
the pairs in $P$ into {\em big pairs}, those with ends in different
microtrees, and {\em small pairs}, those with ends in the same microtree.
Compute path maxima for all the big pairs by running Tarjan's
algorithm on $B$, restricted to the big pairs and using the simple
link-eval structure of Section \ref{sec:link-eval}.

To bound the running time of this computation, we apply Lemma
\ref{lemma:pc} to $B$. Mark every leaf of $B'$.  Each $\ofindroot$
and $\oeval$ occurs in a subtree of $B$ containing a marked node,
so setting $k = 1$ satisfies the hypothesis of the lemma.
Since the number of marked nodes is at most $2n/g =
2n/\log^{1/3}{n}$, the lemma implies an $O(m)$ bound on the time to
compute path maxima for all the big pairs.

We would like to compute path maxima for all the small pairs by
applying the method of Section \ref{sec:tgc}.  To this end,
construct for each microtree a graph 
containing the microtree edges and, for each pair with
both ends in the microtree, an edge designated as a query edge
by a bit.
Now a
new difficulty arises: since the arc costs are arbitrary real
numbers, computing path maxima is not a topological graph
computation; we cannot encode the edge costs in $O(1)$ bits, or even
in $O(\log{g})$ bits.

We overcome this difficulty by following the approach of Dixon,
Rauch, and Tarjan\cite{mst:drt:j}: do a topological graph
computation that builds, for each distinct marked graph, a
comparison tree, whose nodes designate binary comparisons between
costs of unmarked edges of the graph (tree edges), such that the
output nodes of the comparison tree designate, for each marked edge
(query pair), which of the unmarked edges on the path between the
ends of the edge has maximum cost.  Having built all the comparison
trees, run the appropriate comparison tree for each microtree and
its associated pairs, using the actual costs of the microtree arcs
to determine the outcomes of the comparisons.

With $g = \log^{1/3}{n}$, the time for this computation is $O(m)$,
plus the time to build comparison trees for the topologically
distinct instances, plus the time to run the comparison trees for the
actual instances.  Koml\'os \cite{mstver:k} proved that the path maxima
needed for MST verification can be determined in a number of binary
comparisons of tree edge costs
that is linear in the number of graph edges, 
which implies for each instance
the existence of a
comparison tree
that has depth linear in the number of edges. Dixon et
al.~\cite{mst:drt:j} observed that the comparison tree
implied by Koml\'os' result can be built in a time per 
comparison-tree node that is quadratic in the number of graph vertices.
If we use their method to build the comparison trees during the
topological graph computation, then $g = \log^{1/3}{n}$ implies by
the results of Section \ref{sec:tgc} that the total time to build
the comparison trees is $O(m)$.  The total time to run them is
linear in the total size of all the actual instances, which is also
$O(m)$. Thus we obtain:

\begin{theorem}
\label{thm:mst} Computing all the path maxima needed for MST
verification, and doing the verification itself, takes $O(m)$ time
on a pointer machine.
\end{theorem}

\subsection{Construction of Minimum Spanning Trees}

The randomized linear-time MST construction algorithm of Karger,
Klein, and Tarjan~\cite{mst:kkt} runs on a pointer machine except
for the part that computes the path maxima needed for MST
verification. Using the algorithm of Section \ref{sec:mst:path_max},
this part can be done (deterministically) in linear time on a
pointer machine, resulting in a randomized linear-time, pointer
machine algorithm for constructing an MST.

\subsection{Remarks}
\label{sec:mst:remarks}

It is instructive to compare our MST verification algorithm to
those of Dixon, Rauch, and Tarjan~\cite{mst:drt:j} and of
King~\cite{mstver:j:king}. Our use of King's \boruvka~tree
construction as an intermediate step allows us to use only
bottom-level microtrees, whereas Dixon et al.~partition the original
tree entirely into microtrees, with an extra \emph{macrotree} to
represent the connections between them.  It also allows us to use
the simple link-eval structure instead of the sophisticated one.
Lemma \ref{lemma:pc} allows us to break big queries into only two
parts (having an NCA in common); Dixon et al.~break each big query
into as many as six parts.  King explicitly implements Koml\'os'
comparison algorithm for the \boruvka~tree, but her algorithm is
heavily table-driven and requires a RAM.  She also 
must compute NCAs separately.

There is an alternative, though more complicated way to verify an
MST in linear time on a pointer machine.  This method replaces the
use of the \boruvka~tree by a partition of the original tree into
bottom-level microtrees and a set of maximal paths that partition
the core.  The method does NCA computations on trees derived from
the maximal paths, and it uses a sophisticated link-eval structure
instead of the simple one.  We discuss this method in more detail in
Section \ref{sec:alt_step2}.  Though the use of the \boruvka~tree
gives us a simpler algorithm for MST verification, there is no
corresponding concept for either of our remaining applications, and
we must rely on the alternative of partitioning the core into
maximal paths.

\section{Interval Analysis}
\label{sec:ia}

\begin{figure}[t]
\begin{center}
\scalebox{0.825}[0.825]{\input{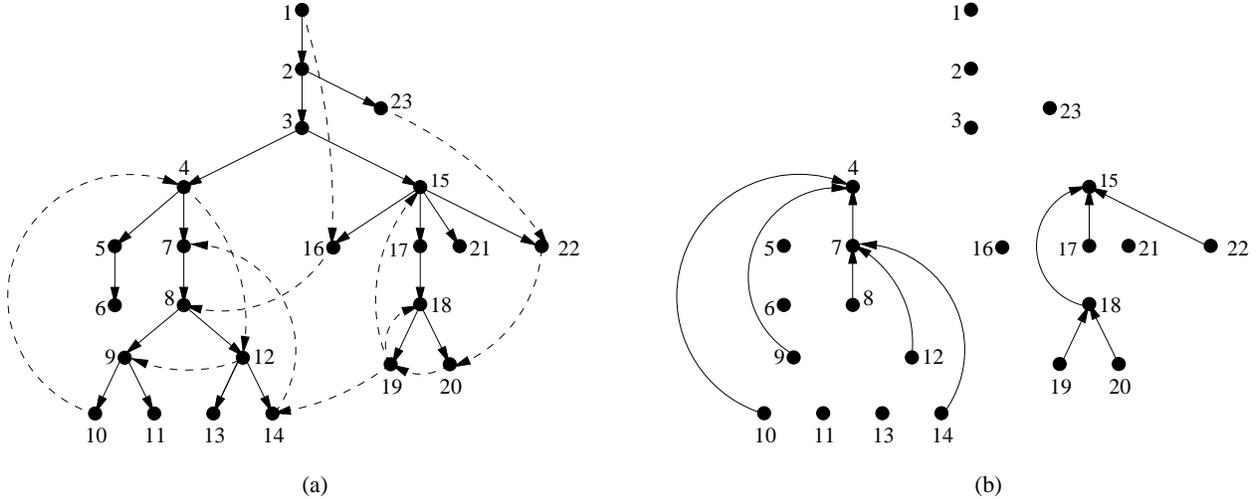}}
\end{center} \caption{(a) A DFS tree $D$
of the input flowgraph $G$; 
non-tree arcs are dashed. (b)
The interval forest $H$ of $G$ with respect to $D$;
arrows are parent pointers.
\label{fig:iforest}}
\end{figure}

We turn now to two problems on flowgraphs.  The first is
\emph{interval analysis}.  Let $G =(V, A, r)$ be a flowgraph, and
let $D$ be a given depth-first search tree rooted at $r$.  Identify
vertices by their preorder number with respect to the DFS: $v < w$
means that $v$ was visited before $w$.  
{\em Reverse preorder} of the vertices
is decreasing order by (preorder) vertex number.
For each vertex $v$, the
\emph{head} of $v$ is
\begin{eqnarray*}
h(v)  =  \max \{ u   : u \neq v \mbox{ and there is a path
from } v  \mbox{ to } u \mbox{ containing only descendants of } u \};
\end{eqnarray*}
$h(v) = \nul$ if this set is empty. The heads define a forest $H$
called the \emph{interval forest}: $h(v)$ is the parent of  $v$ in
$H$. Each subtree $H(v)$ of $H$ induces a strongly connected
subgraph of $G$, containing only vertices in $D(v)$ (the descendants
of $v$ in $D$). See Figure~\ref{fig:iforest}. Tarjan~\cite{st:t}
proposed an algorithm that uses an NCA computation, incremental
backward search, and a DSU data structure to compute $H$ in
$O(m\alpha(m, n))$ time on a pointer machine. We shall add
microtrees, a maximal path partition of the core, and a stack to
Tarjan's algorithm, thereby improving its running time to $O(m)$ on
a pointer machine.

Tarjan's algorithm proceeds as follows.  Delete all the arcs
from the graph.  For each vertex $u$, form a set of all deleted arcs
$(x, y)$ such that $\nca{x, y} = u$. Process the vertices in any
bottom-up order; reverse preorder will do.  To process a vertex $u$,
add back to the graph arcs corresponding to all the deleted arcs
$(x, y)$ with $\nca{x, y} = u$. Then examine each arc $(v, u)$
entering $u$.  If $v \neq u$, set $h(v) = u$, and contract $v$ into
$u$; for all arcs having $v$ as an end, replace $v$ by $u$. This may
create multiple arcs and loops, which poses no difficulty for the
algorithm. Continue until all arcs into $u$ have been examined,
including those formed by contraction.  When adding arcs back to the
graph, the arc corresponding to an original arc is the one formed by
doing end replacements corresponding to all the contractions done so
far.

To keep track of contractions, Tarjan's algorithm uses a DSU
structure whose elements are the graph vertices.  The algorithm also
uses a reverse adjacency set $R(u)$, initially empty, for each
vertex $u$.  A more detailed description of the algorithm is as
follows. To process $u$, for each arc $(x,y)$ such that $\nca{x, y}
= u$, add $x$ to $R(\find(y))$. (The replacement for $x$ is done
later.) Then, while $R(u)$ is non-empty, delete a vertex $x$ from
$R(u)$; let $v \leftarrow \find(x)$; if $v \neq u$, set $h(v)
\leftarrow u$, set $R(u) \leftarrow R(u) \cup R(v)$, and do
$\unite(u,v)$.

With the sets $R(u)$ represented as singly linked circular lists (so
that set union takes constant time), the running time of this
algorithm on a pointer machine is linear except for the NCA
computations and the DSU operations, which take $O(m \alpha(m, n))$
time in Tarjan's original implementation.  We shall reduce the
running time to linear
by using microtrees to eliminate redundant computation
and by reordering the unites into a bottom-up order.

As in Section \ref{sec:nca}, partition $D$ into a set of
bottom-level microtrees (the fringe), each with fewer than $g =
\log^{1/3}{n}$ vertices, and $D'$, the remainder of $D$ (the core).
Use a topological graph computation to compute $h(v)$ for every
vertex $v$ such that $h(v)$ is in the fringe.
The definition of heads implies that for any such vertex
$v$, $h(v)$ and $v$ are in the same microtree,
and furthermore that
the only information
needed to compute heads in the fringe is, for each microtree, the
subgraph induced by its vertices, with non-tree edges marked
by a bit.  With
$g = \log^{1/3}{n}$, this computation takes $O(m)$ time by
Theorem~\ref{thm:tgc2}.

It remains to compute heads for vertices whose heads are in the
core. Our approach is to run Tarjan's algorithm starting from the
state it would have reached after processing the fringe.  This
amounts to contracting all the strong components in the fringe and
then running the algorithm.  This approach does not quite work as
stated, because the DSU operations are not restricted enough for
Lemma \ref{lemma:pc} to apply.  To overcome this difficulty, we
partition the core into maximal paths. Then we run Tarjan's
algorithm path-by-path, keeping track of contractions with a hybrid
structure consisting of a DSU structure that maintains contractions
outside the path being processed and a stack that maintains
contractions inside the path being processed.  The latter structure
functions in the same way as the one Gabow used in his
algorithm\cite{pathdfs:g00} for finding strong components. Now we
give the complete description of our algorithm.

Partition the vertices in $D'$ into a set of maximal paths by
choosing, for each non-leaf vertex $v$ in $D'$, a child $c(v)$ in
$D'$. (Any child will do.)  The arcs $(v, c(v))$ form a set of paths
that partition the vertices in $D'$.  For such a path $P$, we denote
the smallest and largest vertices on $P$ by $\ltop{P}$ and
$\lbottom{P}$, respectively; $\lbottom{P}$ is a leaf of $D'$. Since
$D'$ has at most $n/g$ leaves, the number of paths is at most $n/g$.
Partitioning $D'$ into paths takes $O(n)$ time.

After constructing a maximal path partition of the core, initialize
a DSU structure containing every vertex (fringe and core) as a
singleton set.  Visit the fringe vertices in bottom-up order, and,
for each fringe vertex $v$ with $h(v)$ also in the fringe, perform
$\unite(h(v), v)$; for such a vertex, $h(v)$ has already been
computed. Initialize $R(u) \leftarrow \emptyset$ for every vertex
$u$.  For every arc $(x, y)$ with $x$ and $y$ in the same microtree,
add $x$ to $R(\find(y))$. For every remaining arc $(x, y)$, compute
$u=\nca{x, y}$ and add $(x, y)$ to the set of arcs associated with
$u$. These NCA computations take $O(m)$ time using the algorithm of
Section \ref{sec:nca}. Indeed, every NCA query is big, so the AHU
algorithm answers them in linear time.  \ignore{(The NCA computation does
need a separate DSU structure, whose underlying tree is $D$; the DSU
structure for Tarjan's algorithm has $H$ as the underlying tree.)}
This completes the initialization.

\begin{figure}
\scalebox{0.82}[0.82]{\input{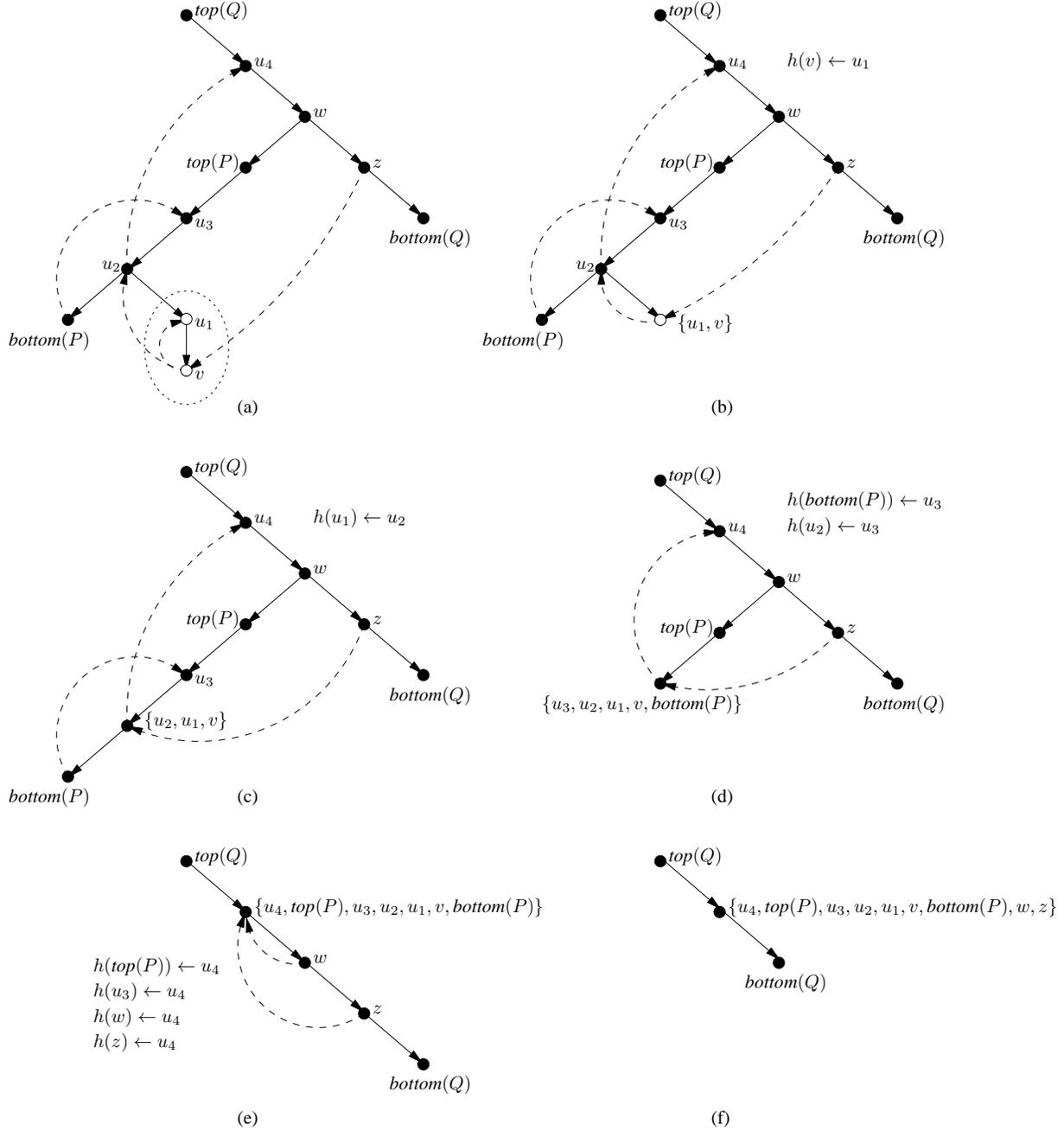}}
\caption{
Idealized execution of the algorithm on the graph in (a),
with circled microtree.
Arcs depict the effects of contractions:
whenever $x\in R(y)$, $(\find(x),\find(y))$ is an arc
in the corresponding graph.
The first vertex in each labeled set is the corresponding original vertex
in (a).
(a$\rightarrow$b) During preprocessing, $h(v)\leftarrow u_1$, and $v$ is inserted
into the set of $u_1$.
(b$\rightarrow$c) When processing $u_2$, $h(u_1)\leftarrow u_2$ via the arc
	$(v,u_2)$.
(c$\rightarrow$d) When processing $u_3$, the stack $S$ is (top-down)
	$(u_2,\lbottom{P})$.  Hence, when processing the arc
	$(\lbottom{P},u_3)$, $S$ is popped so that
	$h(u_2)\leftarrow u_3$ and $h(\lbottom{P})\leftarrow u_3$.
	(d) shows the state
	after doing the $\unite(\cdot)$'s for path $P$.
(d$\rightarrow$e) When processing $u_4$, $S$ is $(w,z,\lbottom{Q})$.
	Arc $(u_2,u_4)$ sets $h(u_3)\leftarrow u_4$ and
	adds $\ltop{P}$ and $z$ to $R(u_4)$.
	Processing $\ltop{P}$ causes
	$h(\ltop{P})\leftarrow u_4$,
	and processing $z$ pops the stack so that 
	$h(w)\leftarrow u_4$ and $h(z)\leftarrow u_4$.
(f) After processing path $Q$.
\label{fig:ia}}
\end{figure}

Now process each path $P$ in the path partition, in bottom-up order
with respect to $\ltop{P}$.  To process a path $P$, initialize 
an empty
stack $S$. Process each vertex $u$ of $P$ in bottom-up
order. To process $u$, for each arc $(x, y)$ such that $\nca{x, y} =
u$, add $x$ to $R(\find(y))$.  Then, while $R(u)$ is non-empty,
delete a vertex $x$ from $R(u)$.  Let $v \leftarrow \find(x)$.  If
$v$ is not on $P$, set $h(v) \leftarrow u$, set $R(u) \leftarrow
R(u) \cup R(v)$, and do $\unite(u, v)$. If, on the other hand, $v$
is on $P$, $v \neq u$, and $v$ is greater than the top vertex on
$S$, pop from $S$ each vertex $w$ less than or equal to $v$, set
$h(w) \leftarrow u$, and set $R(u) \leftarrow R(u) \cup R(w)$. Once
$R(u)$ is empty, push $u$ onto $S$.  After processing all vertices
on $P$, visit each vertex $u$ on $P$ again, in bottom-up order, and
if $h(u)$ is now defined, perform $\unite(h(u), u)$.
See Figure \ref{fig:ia}

This algorithm delays the unites for vertices on a path until the
entire path is processed, using the stack to keep track of the
corresponding contractions.  Specifically, the algorithm maintains
the following invariant: if vertex $u$ on path $P$ is currently
being processed and $x$ is any original vertex, then the vertex into
which $x$ has been contracted is $v = \find(x)$ if $v$ is not on
$P$, or the largest vertex on $S$ less than or equal to $v$ if $v$
is on $P$ and $S$ is non-empty, or $u$ otherwise.  It is
straightforward to verify this invariant by induction on time; the
correctness of this implementation of Tarjan's algorithm follows.

\begin{theorem}
\label{thm:ia} The interval analysis algorithm runs in $O(m)$ time
on a pointer machine.
\end{theorem}

\begin{proof}
The running time is linear except for the find operations: each
vertex gets added to $S$ once and has its head set at most once. To
bound the time for the find operations, we apply Lemma
\ref{lemma:pc} to the tree built by the parent assignments done by
the unite operations. Mark the tops of all paths.  Since there are
at most $n/g$ paths, there are at most $n/g = n/\log^{1/3}{n}$
marked vertices.  We claim that $k = 4$ satisfies the hypothesis of
the lemma.  We need a property of the interval forest $H$: if $h(v)
= u$, then every vertex $w \neq u$ on the path in $D$ from $u$ to
$v$ is a descendant of $u$ in $H$. This holds because there is a path
containing only vertices in $D(u)$ from $w$ to $v$ (via $D$) to $u$.

The unites occur in batches, one initial batch for all the microsets
and one batch per path.  Consider any vertex $v$.  We bound the
number of times the set containing $v$ in the DSU structure can
change, as a result of a batch of unites, before $v$ is in a set
with a marked vertex.  Vertex $v$ can change sets once as a result
of the initialization (from a singleton set to a larger set). After
the initialization, $v$ is in some set, whose designated vertex may
be fringe or core.  The first batch of unites that changes the set
containing $v$ puts $v$ in a set with a designated vertex $u$ that
is in the core, specifically on some path $P$. The second batch of
unites that changes the set containing $v$ puts $v$ in the same set
as $\ltop{P}$ (by the property above), and $v$ is now in a set with
a marked node. Thus $v$ can change sets at most thrice before it is
in a set with a marked vertex.   The parent of $v$ can only change
once, as a result of a compression, without $v$ changing sets.  
Therefore, the parent of $v$ can change at most four times before
$v$ is in a set with a marked vertex, so the claim is true.

With $k = 4$ and $\ell \le n/\log^{1/3}{n}$, Lemma \ref{lemma:pc}
gives a bound of $O(m)$ on the time for the $\find$ operations.
\end{proof}

Interval analysis is an important component of program flow
analysis~\cite{aho:dragon2}. It also has other applications,
including testing flow graph reducibility \cite{reducibility:tarjan},
finding a pair of arc-disjoint spanning trees in a directed
graph \cite{st:t} and verifying a dominator tree \cite{domv:gt05}.
Our interval analysis algorithm gives $O(m)$-time
algorithms on a pointer machine for these applications as well.

In the next section we shall need a compressed version of the
interval forest $H'$ that is defined with respect to the fringe-core
partition: the parent $h'(v)$ of a vertex $v$ is its nearest core
ancestor in $H$ if it has one, $\nul$ otherwise.  We can easily
compute $H'$ from $H$ in linear time, but if we only want $H'$ and
not $H$, we can avoid the topological graph computation on the
microtrees: First, find the strong components of the graphs induced
by the vertex sets of the microtrees. For each such component, find
its smallest vertex $u$, and perform $\unite(u, v)$ for every other
vertex $v$ in the component.  Then run the algorithm above for the
core.  This computes $h(v) = h'(v)$ for every vertex $v$ with head
in the core. Complete the computation by setting $h'(v) = h'(u)$ for
each vertex $v \neq u$ in a fringe strong component with smallest
vertex $u$.

\section{Dominators}
\label{sec:dom}

Our second flowgraph problem is finding immediate dominators.  Let
$G = (V,A,r)$ be a flowgraph.  We denote the immediate dominator of
any vertex $v$ by $\idom{v}$.  Let $D$ be an arbitrary but fixed
depth-first search (DFS) tree rooted at $r$.
As in Section \ref{sec:ia}, we identify
vertices by their preorder number with respect to the DFS;
reverse preorder is decreasing order by vertex number.  We use
the notation $v \anc w$ to denote that $v$ is an ancestor of $w$ in
$D$, and $v \panc w$ to denote that $v$ is a proper ancestor of $w$
in $D$. Sometimes we use the same notation to denote the respective
paths in $D$ from $v$ to $w$. We denote by $p(v)$ the parent of $v$
in $D$. We shall need the following basic property of depth-first
search:

\begin{lemma} \emph{\cite{dfs:t}}
\label{lemma:dfs} Any path from a vertex $v$ to a vertex $w > v$
contains a common ancestor of $v$ and $w$.
\end{lemma}

We shall describe an algorithm to compute immediate dominators in
$O(m)$ time on a pointer machine.  This is our most complicated
application: it uses all the ideas and algorithms we have developed
so far.  Our algorithm is a re-engineering of the algorithms
presented 
by Buchsbaum et al.~\cite{domin:bkrw} and Georgiadis and Tarjan \cite{dom:gt04, Geo05}.
As we proceed with
the description, we shall point out the relationships between
concepts we introduce here and the corresponding ideas in
those previous works.

\subsection{Semi-Dominators, Relative Dominators, Tags, and Extended
Tags}

Lengauer and Tarjan (LT) \cite{domin:lt} devised a three-pass,
$O(m\alpha(m,n))$-time algorithm to
compute immediate dominators.  We shall
improve their algorithm by speeding up the first two steps.  Central
to the LT algorithm is the concept of \emph{semi-dominators}.  A
path $x_0, x_1, \ldots, x_k$ in $G$ is a \emph{high path} if $x_i >
x_k$ for $i < k$. 
As a degenerate case,
a single vertex is a high path.
A high path avoids all proper ancestors of its
last vertex. The \emph{semi-dominator} of a vertex $w$ is
\begin{eqnarray*}
\sdom{w} = \min(\{w\} \cup \{ u : \mbox{for some} \ (u,v) \
\mbox{in} \ A \ \mbox{there is a high path from} \ v \ \mbox{to} \ w
\}).
\end{eqnarray*}
The \emph{relative dominator} of a vertex $w$ is
\begin{eqnarray*}
\rdom{w} = \argmin\{\sdom{u} : \sdom{w} \panc u \anc w\}.
\end{eqnarray*}
With this definition, relative dominators are not unique, but for
any vertex any relative dominator will do.

The LT algorithm operates as follows:

\begin{description}
\item[Step 1:] Compute semi-dominators.
\item[Step 2:] Compute relative dominators from semi-dominators.
\item[Step 3:] Compute immediate dominators from relative dominators.
\end{description}

Step 3 relies on the following lemma:

\begin{lemma}[\protect{\cite[Cor.~1]{domin:lt}}]
\label{lemma:rdom-to-idom} For any vertex $v \neq r$, $\idom{v} =
\sdom{v}$ if $\sdom{\rdom{v}} = \sdom{v}$; otherwise, $\idom{v} =
\idom{\rdom{v}}$.
\end{lemma}

Using Lemma \ref{lemma:rdom-to-idom}, the LT algorithm performs Step
3 in a straightforward top-down pass over $D$ that takes $O(n)$ time
on a pointer machine.

The LT algorithm performs Steps 1 and 2 in a single pass that visits
the vertices of $D$ in reverse preorder and uses a link-eval data
structure to compute semi-dominators and relative dominators.  We
shall present separate algorithms for Steps 1 and 2, although these
steps can be partially combined, as we discuss in Section
\ref{sec:alt_step2}.

Step 2 is almost identical to MST verification.  Indeed, suppose we
assign a cost $\sdom{v}$ to each tree edge $(p(v),v)$ and apply the
MST verification algorithm to the tree $D$ with query set $Q = \{
(sdom(v),v) :  v \neq r \}$, with the modification that the answer
to a query is an edge of minimum cost on the query path rather than
the cost of such an edge. Then for $v \neq r$, $\rdom{v}$ is the
vertex $u$ such that $(p(u),u)$ is the answer to the query
$(\sdom{v},v)$. Modifying the link-eval structure to replace maximum
by minimum and to return edges (or, better, vertices) rather than
costs is straightforward. The algorithm of Section \ref{sec:mst}
thus performs Step 2 in $O(n)$ time on a pointer machine.  (The
number of queries is $O(n)$.)

It remains to implement Step 1, the computation of semi-dominators.
Lengauer and Tarjan reduce this computation, also, to a problem of
finding minima on tree paths, using the following lemma:

\begin{lemma}[\protect{\cite[Thm.~4]{domin:lt}}]
\label{lemma:sdom} 
For any vertex $w$,
$$
\sdom{w} = \min \big(  \{w\} \cup \{ \nca{u,w} : (u,w)\in A \} \cup  
 \{ \sdom{v} : \exists (u,w)\in A,\ \nca{u,w} \panc v \anc u \} \big).
$$
\end{lemma}

The lemma gives a recurrence for $\sdom{w}$ in terms of $\sdom{v}$
for $v
> w$. The LT algorithm performs Step 1 by visiting the vertices in
reverse preorder and using a link-eval structure to perform the
computations needed to evaluate the recurrence.

Even though Step 1 is now reduced to computing minima on tree paths,
we cannot use the MST verification algorithm directly for this
purpose, because that algorithm answers the queries in an order
incompatible with the requirements of the recurrence.  Instead we
develop an alternative strategy.  For convenience we restate the
problem, which allows us to simplify slightly the recurrence in Lemma
\ref{lemma:sdom}. Suppose each vertex $w$ has an integer
\emph{tag} $t(w)$ in the range $[1,n]$.  The \emph{extended tag} of
a vertex $w$ is defined to be
\begin{eqnarray*}
\et{w} = \min \{ t(v) : \mbox{there is a high path from} \ v \
\mbox{to} \ w \}.
\end{eqnarray*}

\begin{lemma}
\label{lemma:et} If $t(w) = \min ( \{ w \} \cup \{ v : (v,w)\in A\} )$ for every vertex, then $\sdom{w} = \et{w}$ for
every vertex.
\end{lemma}
\begin{proof}
Immediate from the definitions of semi-dominators and extended tags.
\end{proof}

We can easily compute the tag specified in Lemma \ref{lemma:et} for
every vertex in $O(m)$ time.  Thus the problem of computing
semi-dominators becomes that of computing extended tags.

Lemma \ref{lemma:sdom} extends to give the following recurrence for
extended tags:

\begin{lemma}
\label{lemma:et2}
For any vertex $w$,
\begin{eqnarray*}
\et{w} = \min(\{t(w)\} \cup \{ \et{v} : \exists (u,w) \in A,\ \nca{u,w} \panc v \anc u
\}).
\end{eqnarray*}
\end{lemma}
\begin{proof}
Analogous to the proof of Lemma \ref{lemma:sdom}.  Let $x$ be the
right side of the equation in the statement of the lemma.  First we
prove $\et{w} \le x$. If $x = t(w)$, $\et{w} \le x$ is immediate
from the definition of $\et{w}$. Suppose $x = \et{v}$ for $v$ such
that $\nca{u,w} \panc v \anc u$ and $(u,w)$ in $A$.  By the
definition of $\et{v}$, $\et{v} = t(z)$ for some vertex $z$ such
that there is a high path from $z$ to $v$. Extending this path by
the tree path from $v$ to $u$ followed by the arc $(u,w)$ gives a
high path from $z$ to $w$. Hence $\et{w} \le \et{v} = x$.

Next we prove $x \le \et{w}$.  Let $z$  be a vertex such that
$\et{w} = t(z)$ and there is a high path from $z$ to $w$ (by the
definition of the extended tags).  If $z =
w$, then $x \le \et{w}$ from the definition of $x$.  If not, let
$(u,w)$ be the last edge on the high path from $z$ to $w$.  Let $v$
be the first vertex along the high path such that $\nca{u,w} \panc v
\anc u$. Such a $v$ exists since $u$ is a candidate ($\nca{u,w} \le
w < u$). We claim that the part of the high path from $z$ to $v$ is
itself a high path. Suppose to the contrary that this part contains
a vertex less than $v$, and let $y$ be the last such vertex.  Then
$y$ must be an ancestor of $v$ by Lemma \ref{lemma:dfs},
and since $y$ is on a high path for $w$, $\nca{u,w} \panc y \panc v$.
This
contradicts the choice of $v$. It follows that  $\et{v} \le t(z)$;
that is, $x \le \et{w}$.
\end{proof}

\begin{figure}[t]
\begin{center}
\scalebox{1}[1]{\input{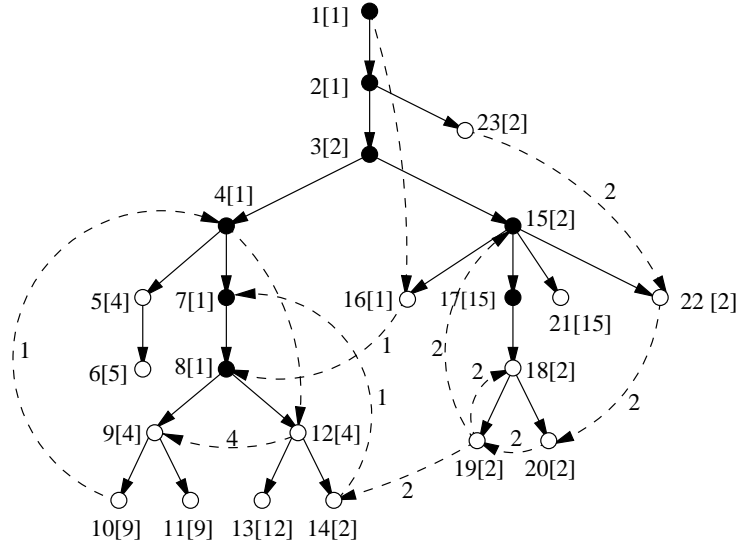}}
\end{center} \caption{
Extended tags and arc tags. The number inside
each bracket is the extended tag of
the corresponding vertex. The number on each arc
is its tag; the arc tag of a tree or forward arc
is infinite and not shown in the figure.
\label{fig:tags}}
\end{figure}

We introduce one more definition that simplifies some of our
formulas and discussion.  For an arc $(u,w)$, the \emph{arc tag} of
$(u,w)$ is
\begin{eqnarray*}
\at{u,w} = \min \{ \et{v} : \nca{u,w} \panc v \anc u
\}
\end{eqnarray*}
if this minimum is over a non-empty set, and infinity otherwise
(when $\nca{u,w} = u$). An example is shown in Figure \ref{fig:tags}.
Using arc tags, the recurrence in Lemma
\ref{lemma:et2} becomes
\begin{eqnarray}
\label{eq:et} \et{w} = \min(\{ t(w)\} \cup \{ \at{u,w} : (u,w)\in A\}).
\end{eqnarray}

\subsection{The Interval Forest}

We could use \eqref{eq:et} to compute extended tags just as the LT
algorithm uses Lemma \ref{lemma:sdom} to compute semi-dominators,
but we seek a faster method. Note
that there are two kinds of arcs $(u,w)$ that must be handled: those
such that $u$ and $w$ are unrelated (\emph{cross arcs}), and those
such that $w \panc u$ (\emph{back arcs}).  (Arcs such that $u \panc
w$ do not contribute to the recurrence.)    We apply different
techniques to the cross arcs and the back arcs, which allows us to
tease apart the intertwined computations implied by \eqref{eq:et}
and reorder them to apply our techniques.

To handle the back arcs, we use the interval forest discussed in
Section \ref{sec:ia}.  Recall the following definitions.  For each
vertex $w$, the head $h(w)$ of $w$ is the maximum vertex $u \neq w$
such that there is a path from $w$ to $u$ containing only
descendants of $u$, if this maximum is over a non-empty set, and
$\nul$ otherwise. Lemma \ref{lemma:dfs} implies that the constraint
on $u$ in the definition of $h(w)$ is equivalent to $u \panc w$ and
there is a high path from $w$ to $u$. The heads define a forest $H$
called the \emph{interval forest}: $h(w)$ is the parent of $w$ in
$H$. The following lemma allows us to compute extended tags by
computing arc tags only for the cross arcs and propagating minima up
the interval forest.

\begin{lemma}
\label{lemma:at-to-et} For any vertex $w$,
\begin{eqnarray*}
\et{w} = \min ( \{ t(v) : v \in H(w) \} \cup \{
\at{u,v} : (u,v) \in A,\ v \in H(w),\ u \not\in D(w) \} ).
\end{eqnarray*}
\end{lemma}
\begin{proof}
Let $x$ be the right side of the equation in the statement of the
lemma.  First we prove $\et{w} \le x$.  Let $v$ be in $H(w)$.  Since
there is a high path from $v$ to $w$, $\et{w} \le t(v)$.  Let
$(u,v)$ be in $A$ such that $v$ is in $H(w)$ but $u$ is not in
$D(w)$. Let $y$ be a vertex of minimum $\et{y}$ such that $\nca{u,v}
\panc y \anc u$, and let $z$ be a vertex of minimum $t(z)$ such that
there is a high path from $z$ to $y$. Then there is a high path from
$z$ to $y$ to $u$ to $v$ to $w$, which implies $\et{w} \le t(z) =
\et{y} = \at{u,v}$. We conclude that $\et{w} \le x$.

Next we prove $x \le \et{w}$.  Let $z$ be a vertex such that $\et{w}
= t(z)$ and there is a high path from $z$ to $w$.  If $z$ is in
$H(w)$, then $x \le t(z) = \et{w}$.  Suppose $z$ is not in $H(w)$.
Let $(u,v)$ be the first arc along the high path from $z$ to $w$
such that $v$ is in $H(w)$. Then $u$ cannot be in $D(w)$, or it
would be in $H(w)$, contradicting the choice of $(u,v)$.  
Thus $\nca{u,v}\panc u$.
Let $y$ by the first vertex along the high path such that 
$\nca{u,v}\panc y\anc u$.
By Lemma
\ref{lemma:dfs}, the part of the high path from $z$ to $y$ is itself
a high path. Thus $x \le \at{u,v} \le \et{y} \le t(z) = \et{w}$.
\end{proof}

\begin{corollary}
\label{cor:et} For any vertex $w$,
\begin{eqnarray*}
\et{w} = \min ( \{ t(w) \} \cup \{ \et{v} : h(v) = w \} \cup \{
\at{v,w} : (v,w) \ \mbox{\rm is a cross arc} \}).
\end{eqnarray*}
\end{corollary}

Corollary \ref{cor:et} gives an alternative recursion for
computing extended tags by processing the vertices in reverse
preorder. Lemma \ref{lemma:at-to-et} also allows us to compute
extended tags for all the vertices on a tree path, given only arc
tags for arcs starting to the right of the path.

\subsection{Microtrees and Left Paths}

As in Section \ref{sec:nca}, we partition $D$ into a set of
bottom-level microtrees (the fringe), each containing fewer than $g =
\log^{1/3}{n}$ vertices, and $D'$ (the core), the remainder of $D$.
We call a cross arc \emph{small} if both its ends are in the same
microtree and \emph{big} otherwise.  We also partition $D'$ into
maximal paths as in Section \ref{sec:ia}, but a particular set of
maximal paths. Specifically, we partition $D'$ into \emph{left
paths}, as follows: an arc $(p(v),v)$ of $D'$ is a \emph{left arc}
if $v$ is the smallest child of $p(v)$ in $D'$.  A \emph{left path}
is a maximal sequence of left arcs.  We can partition $D$ into
microtrees and left paths in $O(m)$ time during the DFS that defines
$D$.  If $P$ is a left path, as in Section \ref{sec:ia} we denote by
$\ltop{P}$ and $\lbottom{P}$ the smallest and largest vertices on
$P$, respectively. The importance of left paths is twofold. First,
there are at most $n/g$ of them. Second, if $(p(v),v)$ is a left
arc, any child of $p(v)$ smaller than $v$ must be in the fringe, not
the core. That is, left paths have only microtrees descending on
their left. Left paths serve in place of the \emph{lines} of
Georgiadis and Tarjan
\cite{dom:gt04,Geo05}; left paths are catenations of those lines.

Our hypothetical plan for computing extended tags in linear time is
to use a topological graph computation to handle the microtrees and
a link-eval structure to compute arc tags for the big cross edges.
This plan does not quite work: computing extended tags is unlike the
previous problems we have considered in that there is an interaction
between the fringe and the core.  In particular, we need at least
some information about the small cross arcs in order to compute
extended tags in the core, and information about the big cross arcs
to compute extended tags in the fringe.  For the former computation
we do not, however, need to compute arc tags for the small cross
arcs: the recurrence in Lemma \ref{lemma:at-to-et} expresses the
extended tags of vertices in the core in terms only of tags of
vertices and arc tags of big cross arcs.  To handle the limited
interaction between fringe and core, we use a two-pass strategy.
During the first pass, we compute arc tags of big cross arcs and
extended tags in the core while computing limited information in the
fringe.  In the second pass, we use the information computed in the
first pass in a topological graph computation to compute extended
tags in the fringe.

The information we need in the fringe is a set of values defined as
follows.  For a vertex $w$  in a microtree $D(s)$, the
\emph{microtag} of $w$ is
\[
\begin{split}
\mt{w} = & \min \big(  \{ t(v) : \mbox{there is a path from} 
\ v \ \mbox{to} \ w \ \mbox{in} \ D(s) \} \ \cup \\
& \{ \at{u,v} : (u,v) \ \mbox{is a cross arc}, \ v  \in D(s), \ u \notin D(s), 
\ \mbox{and there is a path in} \ D(s) \ \mbox{from} \ v \ \mbox{to} \
w \} \big). \\
\end{split}
\]
Our microtags correspond to the \emph{pushed external dominators} of
Buchsbaum et al.~\cite{domin:bkrw} (also used 
by Georgiadis and Tarjan \cite{dom:gt04,Geo05}). The next
lemma shows that, when computing the arc tags of big cross arcs, we
can use microtags in place of extended tags for fringe vertices;
that is, we shall use microtag values in the link-eval structure, when
linking fringe vertices.

\begin{figure}[t]
\begin{center}
\scalebox{1}[1]{\input{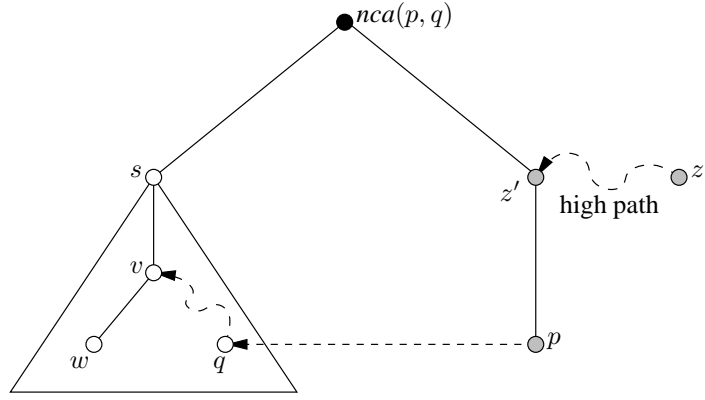}}
\end{center} \caption{Proof of Lemma \ref{lemma:mt-et}.
Dashed curves represent graph paths. 
Solid edges represent tree paths.
Each gray vertex can be in the core or in the fringe.
\label{fig:mt-et}}
\end{figure}

\begin{lemma}
\label{lemma:mt-et} Let $w$ be a vertex in a microtree $D(s)$.  Then
\begin{eqnarray*}
\min \{ \et{v} : s \anc v \anc w \} = \min \{ \mt{v} : s
\anc  v \anc w \}.
\end{eqnarray*}
\end{lemma}
\begin{proof} Let $x$ and $y$ be the values of the left and right sides
of the equation in the statement of the lemma, respectively.  First
we prove that $x \ge y$.  Let $v$ be a vertex such that $x = \et{v}$
and $s \anc v \anc w$.  Let $z$ be a vertex such that $t(z) =
\et{v}$ and there is a high path from $z$ to $v$.  If $z$ is in
$D(s)$, then this high path is in $D(s)$, which implies that $x =
t(z) \ge mt(v) \ge y$. Suppose on the other hand that $z$ is not in
$D(s)$. Let $(p,q)$ be the last arc along the high path such that
$p$ is not in $D(s)$, and let $z'$ be the first vertex along the high
path such that $\nca{p,q} \panc z' \anc p$.  Note that $(p,q)$ must
be a cross arc, since $p$ is not in $D(s)$ and is on a high path to
$v$ in $D(s)$. See Figure \ref{fig:mt-et}.
As in the proof of Lemma \ref{lemma:et2}, the part of
the high path from $z$ to $z'$ is itself a high path, which implies
$x = t(z) \ge \et{z'} \ge \at{p,q} \ge \mt{v} \ge y$.

Next we prove that $x \le y$.  Let $v$ be a vertex such that $y =
\mt{v}$ and $s \anc  v \anc w$.  Suppose $\mt{v} = t(z)$ for some
$z$ in $D(s)$ from which there is a path to $v$ in $D(s)$.  Let $u$
be the first vertex on this path that is an ancestor of $w$.  Then
the path from $z$ to $u$ is a high path by
Lemma \ref{lemma:dfs} and the choice of $u$. Thus $x \le \et{u} \le
t(z) = y$. Suppose on the other hand that $\mt{v} = \at{p,q}$ for an
arc $(p,q)$ such that $q$ but not $p$ is in $D(s)$ and there is a
path from $q$ to $v$. Let $u$ be the first vertex on this path that
is an ancestor of $w$. By Lemma \ref{lemma:dfs}, the part of the
path from $q$ to $u$ is a high path. Let $z$ be a vertex such that
$t(z) = \at{p,q}$ and there is a high path from $z$ to a vertex $z'$ such that
$\nca{p,q} \panc z' \anc p$. See Figure \ref{fig:mt-et}.
This path, together with the path $z' \anc p$, 
the arc $(p,q)$, and the high path
from $q$ to $u$, is a high path.  Thus $x \le \et{u} \le t(z) =
\at{p,q} = \mt{v} = y$.
\end{proof}

To help compute extended tags during the first pass, we use a
compressed interval forest $H'$ in place of the interval forest $H$.  
Recall that in $H'$, the parent $h'(v)$
of a vertex $v$ is the nearest ancestor of $v$ in $H$ that is a core
vertex. Forests $H$ and $H'$ are identical on the core; each subtree
of $H$ consisting of fringe vertices with a core root is compressed
in $H'$ to the root with all the fringe vertices as children.  The
use of $H'$ in place of $H$ is an optimization only: we can build
either $H$ or $H'$ in linear time using the algorithm of Section \ref{sec:ia},
but, as noted in Section \ref{sec:ia}, building $H'$ instead of $H$
avoids the use of topological graph computations on the microtrees
and thus is simpler. The algorithm of Section \ref{sec:ia} builds
$H'$ by partitioning $D$ into microtrees and maximal paths.  We can
use the set of left paths as the maximal paths, avoiding the need
for two different partitions.

To compute extended tags in the core, we use the following
corollary of Lemma \ref{lemma:at-to-et}:

\begin{corollary}
\label{cor:at-to-et} If $w$ is a core vertex
\[
\begin{split}
\et{w} = \min \big( 
& \{ t(v) : v = w \ \mbox{\rm or} \ v
\ \mbox{\rm is fringe with} \ h'(v) = w \} \cup \{ et(v) : v
\ \mbox{\rm is core with} \ h'(v) = w \} \ \cup \\
& \{ \at{u,v} :
(u,v) \ \mbox{\rm is a big cross arc such that} \ v = w \ \mbox{\rm or} \ v
\ \mbox{\rm is fringe with} \ h'(v) = w \} \big ). \\
\end{split}
\]
\end{corollary}

The algorithm of Georgiadis and Tarjan
\cite{dom:gt04} for computing dominators does not
use $H'$ explicitly, but it does do an incremental backward search
using a stack to maintain strongly connected parts of lines, in
effect doing a just-in-time computation of (part of) $H'$.  Making
this computation separate, as we have done, breaks the overall
algorithm into smaller, easier-to-understand parts, which could be
combined if desired.

\subsection{Computation of Arc Tags}
\label{sec:at}

The heart of the algorithm is the computation of arc tags.  We split
each such computation into two parts, either of which can be void: a
\emph{top part}, which computes a minimum of extended tags over part
or all of a left path, and a \emph{bottom part}, which computes a
minimum of extended tags of core vertices and microtags of fringe
vertices using a sophisticated link-eval structure.  Specifically,
let $(u,v)$ be a big cross arc.  Let $P$ be the left path containing
$\nca{u,v}$, and let $Q$ be the intersection of $P$ and the path
$\nca{u,v} \panc u$. 
We denote the last vertex on $Q$ by
$\mv{u,v}$. Note that $Q$ can be non-empty (contain no arcs) only if
$v$ is a fringe vertex. See Figure \ref{fig:mid}.

\begin{figure}[t]
\begin{center}
\scalebox{1}[1]{\input{mid.pstex_t}}
\end{center} \caption{Examples of non-empty $\nca{u,v} \panc \mv{u,v}$ paths.
(a) Case $u>\lbottom{P}$. (b) Case $u<\lbottom{P}$.
\label{fig:mid}}
\end{figure}

For a given left path $P$, we compute minima of extended tags for
all such non-empty paths $Q$ at the same time.  We do not need to
know any of these minima until all the extended tags for
vertices on $P$ have been computed.  This allows us to compute the
minima for such paths $Q$ in arbitrary order.  One way to compute
these minima is to use the MST verification algorithm, as suggested
above for doing Step 2 of the LT algorithm.  In this application,
however, the tree being verified is actually a path, and we can use
an algorithm that is at least conceptually simpler, if not
asymptotically faster.  The problem we need to solve is that of
computing minima for given subsequences of a sequence of numbers.
This is the \emph{range minimum query} (RMQ) problem
\cite{gp:gbt84}. This problem has a linear-time reduction
\cite{gp:gbt84} to an NCA problem on a Cartesian
tree \cite{carttree:vuil}. We can thus compute minima for paths $Q$
by constructing the Cartesian tree and applying our NCA algorithm.
Either method allows us to compute the top parts of arc
tags in $O(m)$ time on a pointer machine.

To compute the bottom parts of arc tags, we use a sophisticated
link-eval structure.  We delay the links for arcs on a left path
until the top of the left path is reached, and for arcs in a
microtree until its root is reached.  This allows us to establish a
linear time bound for all the link-eval operations using Lemma
\ref{lemma:pc}.

\subsection{The First Pass}

We now have all the pieces necessary to describe the first pass of our
algorithm for computing extended tags.  Before the first pass, build
the compressed interval forest $H'$, compute $\nca{u,v}$ for each
big cross arc $(u,v)$, and construct, for each core vertex $w$, the
set of big cross arcs $(u,v)$ with $\nca{u,v} = w$.  This takes
$O(m)$ time on a pointer machine using the method of Section 6: the
NCAs are computed as part of the algorithm that builds $H'$.  Each
vertex $v$ has a \emph{computed tag} $\ct{v}$ that is initialized to
$t(v)$ and that decreases as the first pass proceeds, until $\ct{v}
= \mt{v}$ if $v$ is fringe, or $\ct{v} = \et{v}$ if $v$ is core.  Each
fringe vertex $v$ also has an associated set of cross arcs,
initially empty. For each fringe vertex $v$, if $v$ has a parent in
$H'$ and $\ct{h'(v)} > \ct{v}$, replace $\ct{h'(v)}$ by $\ct{v}$.
Finally, initialize a sophisticated link-eval data structure with no
edges and each vertex of $G$ as a node.

The first pass visits each microtree once and each left path twice.
The visits are in reverse preorder with respect to the roots of the
microtrees and the top and bottom vertices of the left paths;
the first visit to a left path corresponds to its bottom (largest)
vertex;
the second visit, to its top (smallest) vertex.
Conceptually, envision a reverse preorder traversal of $D$,
with actions taken as described below whenever a microtree root
or bottom or top vertex of a left path is visited.

When visiting a microtree $D(s)$, it will be true that, for each
vertex $v$ in $D(s)$,
\begin{equation}
\label{eq:et-ct}
\et{v} \le \ct{v} \le \min(\{t(v)\} \cup 
\{\at{u,v} : (u,v)\in A,\ u \not\in D(s) \}).
\end{equation}
Compute microtags for all vertices in $D(s)$ by finding the strong
components of the subgraph induced by the vertices in $D(s)$ and
processing the strong components in topological order.  To process a
component, compute a microtag for the component, equal to the
minimum of the $\ct{\cdot}$ values for all vertices in the
component and the microtags for all preceding components (those with
an arc leading to the component). Then set $\ct{v}$ for every vertex
in the component equal to the computed microtag.  
The assigned value of $\ct{v}$ must be $\mt{v}$, assuming 
\eqref{eq:et-ct} holds. The time required
for this computation is linear in the size of the subgraph induced
by $D(s)$ \cite{dfs:t}. Having computed microtags for $D(s)$,
perform $\mlink{p(v), v, \ct{v}}$ for every vertex in $D(s)$, in
bottom-up order. Finally, for each cross arc $(u,v)$ in the set of
cross arcs of a vertex $u$ in $D(s)$, set $\ct{v} \leftarrow \min\{\ct{v},
\meval{u}\}$, and then set $\ct{h'(v)} \leftarrow \min\{\ct{h'(v)}, \ct{v}\}$
if $v$ has a parent in $H'$. 
Such computations happen here only for
arcs $(u,v)$ such that $u$ is in a microtree hanging on the left of
some left path. It will become clear later that, for such an arc, 
the top part of the evaluation of
$\at{u,v}$ gets done first, when the left path is processed. The
$\meval{u}$ operation does the bottom part of the evaluation,
finishing the job. We describe below when these 
arcs are entered in the set associated with $u$.

When visiting a left path $P$ for the first time, begin by visiting
the vertices $w$ of $P$ in bottom-up order and setting $\ct{h'(w)} \leftarrow
\min\{\ct{h'(w)}, \ct{w}\}$ if $w$ has a parent in $H'$.  Once these
updates are completed, $\ct{w} = \et{w}$ for every vertex $w$ on
$P$. Then collect all the arcs $(u,v)$ in the sets associated with
the vertices on $P$; i.e., the arcs $(u,v)$ such that $\nca{u,v} \in P$. 
For each such arc $(u,v)$, set $\mv{u,v} \leftarrow
p(\mroot{\micro{u}})$ if $u < \lbottom{P}$, and $\mv{u,v} \leftarrow \findroot{u}$
otherwise. The $\ofindroot$ operation in the latter case is an
operation on the link-eval structure.  Having computed all the
$\mathit{mid}$ values for all the cross arcs, evaluate the top parts
of their arc tags, using either of the methods discussed in Section
\ref{sec:at}.  For each such arc $(u,v)$ with computed arc tag top
part $x$, do the following.  If $u > \lbottom{v}$ (see Figure \ref{fig:mid}a), 
set $x \leftarrow \min\{x, \meval{u}\}$; otherwise (see Figure \ref{fig:mid}b), 
add $(u,v)$ to the set of cross arcs of
$u$. In the former case, the $\meval{u}$ operation computes the
bottom part of the arc tag; in the latter case, the computation of
the bottom part is done when the microtree containing $u$ (which
hangs to the left of $P$) is visited. In either case, set $\ct{v} \leftarrow
\min\{\ct{v}, x\}$, and then set $\ct{h'(v)} \leftarrow \min\{\ct{h'(v)},
\ct{v}\}$ if $v$ is a fringe vertex with a parent in $H'$.

When visiting a left path $P$ for the second time, perform
$\mlink{p(w), w, \ct{w}}$ for each vertex on $P$ in bottom-up order,
unless $P$ is the last path, in which case the first pass is done.

Based on the results of the previous sections, it is straightforward
(but tedious) to prove that this algorithm correctly computes extended
tags. Note that the algorithm eagerly pushes $\ct{\cdot}$ values up
$H'$, rather than lazily pulling them; the latter would require
computing sets of children for $H'$, whereas the former can be done
using just parent pointers.

\begin{lemma}
The first pass takes $O(m)$ time on a pointer machine.
\end{lemma}
\begin{proof}
The running time of all parts of the algorithm is linear based on
previous results, except for the $\ofindroot$ and $\oeval$
operations. To bound the time for these, we apply Lemma
\ref{lemma:pc} to the shadow subtrees built by the link operations.
These subtrees are $\sqrt{2}$-balanced by Corollary
\ref{cor:link-by-size} for linking-by-size and Corollary
\ref{cor:link-by-rank} for linking-by-rank.  Mark the parents (in
$D$) of the tops of all the left paths.  This marks at most $n/g =
n/\log^{1/3}{n}$ vertices. We claim that $k = 5$ satisfies the
hypothesis of the lemma. 

We need to use details of the link
implementation, for which we refer the reader to Section
\ref{sec:link-by-size} for linking-by-size and
\ref{sec:link-by-rank} for linking-by-rank. The links occur in
batches with no intermixed $\ofindroot$ or $\oeval$ operations, one
batch per microtree and one batch per left path. Let $v$ be any
vertex. We count the number of times the subroot of the shadow
subtree containing $v$ can change, as the result of a batch of
links, before $v$ is in a subtree containing a marked node. Let $v_0
= v, v_1, v_2,\ldots$ be the successive roots of the shadow trees
containing $v$. The subroot of the shadow subtree containing $v$ can
change only as the result of a batch of links that include the
current $v_i$ as one of the vertices being linked. Suppose $v$ is
fringe. The first batch of links to include $v_0$ is the one for
$\micro{v}$. This batch of links makes $p(\mroot{\micro{v}})$
the root of the tree containing
$v$; that is, $v_1 = p(\mroot{\micro{v}})$. The next batches of
links that include $v_1$ are those for other microtrees whose roots
are children of $v_1$ in $D$. Such a batch does not change the root
of the tree containing $v$ but can change the subroot of the
subtree containing $v$, making it equal to $v_1$.  Once such links
are done, the only remaining batch of links that includes $v_1$ is
the one for the left path $P_1$ containing $v_1$. This batch makes
$v_2 = p(\ltop{P_1})$, which means that the shadow tree containing
$v$ (but not necessarily the shadow subtree containing $v$) has a
marked vertex. The next batches of links that include $v_2$ are
those for microtrees whose roots are children of $v_2$ in $D$. Such
a batch cannot change the root of the tree containing $v$, but it can
change the subroot of the subtree containing $v$, making it equal to
$v_2$, which is marked. Otherwise, the next (and last) batch of
links that includes $v_2$ is the one for the left path $P_2$
containing $v_2$. This batch makes $v_3 = p(\ltop{P_2})$.  

Now $v$
is either in the subtree rooted at $v_3$, and hence in a subtree
with a marked vertex, or it is a shadow descendant of $v_2$, which
is no longer the root of the shadow tree containing $v$.  No
subsequent link can change the root of the subtree containing $v$
without putting $v$ and $v_2$, which is marked, in the same subtree.
Tracing through the analysis above, we see that the subroot of the
shadow subtree containing a fringe vertex $v$ can change at most
four times before $v$ is in a subtree with a marked vertex. If $v$
is a core vertex, the last part of the same analysis applies: the
first batch of links that can change either the root of the tree
containing $v$ or the subroot of the subtree containing $v$ is the
one for the left path containing $v$; the subroot of the subtree
containing $v$ can change at most twice before $v$ is in a subtree
with a marked vertex. The shadow parent of vertex $v$ can change at
most once before the root of the shadow subtree containing $v$
changes. Thus the shadow parent of $v$ can change at most five times
before $v$ is in a shadow subtree with a marked vertex.  This
verifies the claim.  With $k = 5$ and $\ell \le n/\log^{1/3}{n}$,
Lemma \ref{lemma:pc} gives a bound of $O(m)$ on the time for the
$\ofindroot$ and $\oeval$ operations.
\end{proof}

\subsection{The Second Pass}

Having computed extended tags for all core vertices, we compute
extended tags for all fringe vertices by using a topological graph
computation on the microtrees. In the first pass, just before a
microtree $D(s)$ is processed, each vertex $v$ in $D(s)$ has $\ct{v}
= \min(\{ t(v) \} \cup \{ \at{u,v} : (u,v) \in A,\ u \not\in D(s) \})$. 
It follows that if we
compute extended tags within the subgraph induced by the vertices of
$D(s)$, using these $\ct{\cdot}$ values as the initial tags, we
will obtain the correct extended tags for the vertices in $D(s)$
with respect to the original tags in the entire graph.  The
$\ct{\cdot}$ values are in the range $[1,n]$, but we can map them
to the range $[1,g]$ by sorting all the $\ct{\cdot}$ values using a
pointer-based radix sort, extracting a sorted list of
$\ct{\cdot}$ values for each subproblem, and mapping each such
sorted list to $[1,g]$.  To do this on a pointer machine, we need to
maintain a singly linked master list of length $n$, whose nodes
correspond to the integers 1 through $n$, and store with each
integer a pointer to its corresponding position in the master list,
and we need to track such pointers through the entire running of the
algorithm.  We assume that each input tag is given along with a
corresponding pointer into the master list.  For the special case of
computing semi-dominators, we construct the master list and the
corresponding pointers as we perform the depth-first search and
number the vertices.  The only manipulations of vertex numbers are
comparisons, so it is easy to track these pointers through the
entire computation.

Once the tags are mapped to $[1,g]$, the computation of extended
tags on the microtrees is a topological graph computation, which we
perform using the method described in Section \ref{sec:nca}.  With
the choice $g = \log^{1/3}{n}$, the second pass requires $O(m)$ time
on a pointer machine.

Combining all the parts of the algorithm, we obtain the following
theorem:

\begin{theorem}
\label{thm:dom} Finding immediate dominators takes $O(m)$ time on a
pointer machine.
\end{theorem}

\subsection{An Alternative Method for Step 2}
\label{sec:alt_step2}

We conclude our discussion of dominators by sketching an alternative
method for performing Step 2 (computing relative dominators) that does some of 
the work in the second pass
of Step 1 and then uses a simplification of the algorithm for the first
pass of Step 1 to do the rest.

For a microtree $D(s)$, the $\ct{\cdot}$ values of its vertices
just before $D(s)$ is processed provide enough information not only
to compute the semi-dominators of each of its vertices but also to
compute the relative dominator of each vertex $v$ such that
$\sdom{v}$ is in $D(s)$.  This we can do as part of the topological
graph computation that forms the second pass of Step 1.  The
remaining part of Step 2 is to compute 
$\rdom{v} = \argmin \{ \sdom{u} : \sdom{v} \panc u \anc v \}$ 
for each vertex $v$ with $\sdom{v}$ in
the core. We can do this by running a simplified version of the
first pass of Step 1. We modify the link-eval structure so that an
$\oeval$ returns a vertex of minimum value, rather than the value
itself.  We compute the relative dominators in the same way that
pass 1 of Step 1 computes the arc tags of big cross arcs, but
without using the interval tree $H'$ and without using nearest
common ancestors.  We begin by storing each pair $(\sdom{v},v)$ with
$\sdom{v}$.  Then we perform $\mlink{p(v), v, \sdom{v}}$ for every
fringe vertex $v$, in reverse preorder.  Finally, we process each
left path $P$, in reverse preorder with respect to $\lbottom{P}$. To
process a left path $P$, we collect all the pairs $(u,v)$ stored
with its vertices. For each such pair, we set $\mv{u,v} \leftarrow
\findroot{v}$.  We evaluate each top part from $u$ to $\mv{u,v}$ using an
NCA computation on a derived Cartesian tree as discussed in Section
\ref{sec:at}, modified to return a candidate relative dominator
$\rd{u,v}$ for each pair.  For each pair we set $\rdom{v} \leftarrow
\argmin\{\sdom{\meval{v}}, \sdom{\rd{u,v}}\}$. Finally, we perform
$\mlink{p(v), v, \sdom{v}}$ for every vertex on $P$ in reverse
preorder, unless $P$ is the last path, in which case we are done.
This method for doing Step 2 takes $O(n)$ time.

This approach also leads to an alternative algorithm for MST
verification, as mentioned in Section \ref{sec:mst:remarks}, which
avoids the use of the \boruvka~tree as an intermediate step,
replacing it with NCA computations on Cartesian trees derived from
the paths of a partition of the core of the original tree $T$ into
maximal paths. We must still do verification within microtrees, but
these are microtrees of the original tree rather than of the
\boruvka~tree.

\subsection{Remarks}

From the definition of microtags we have that for any 
$w$ in a microtree $D(s)$, $\mt{w} \le \mt{v}$ for
any $s \anc v \anc w$. This inequality implies that the eval
function need only operate on the core tree. The algorithms
of Buchsbaum et al.~\cite{domin:bkrw} and Georgiadis and Tarjan \cite{dom:gt04, Geo05} rely on this fact
but also require a hybrid link-eval structure for the evaluation
of path minima on the core.
Lemma \ref{lemma:pc} allows us 
to use a standard (simpler) link-eval structure that can include the fringe,
which also yields a more uniform treatment of the core and fringe vertices. 
 
Our dominators algorithm uses the 
linear-time offline NCA algorithm
for two subproblems: interval analysis and range minimum queries.
Georgiadis \cite{Geo05} observed that a refined partition of the core
tree into unary paths of size $O(g)$ enables us to use trivial algorithms to
compute NCAs; topological graph computations are still required, but they 
are performed on Cartesian trees corresponding to each unary path. 

\section{Component Trees}
\label{sec:ct}

Our final application is a tree problem, unusual in that it seems to
require partitioning all of the given tree, rather than
just the bottom part, into microtrees.

\subsection{Kruskal Trees}

\begin{figure}[t]
\begin{center}
\scalebox{.95}[.95]{\input{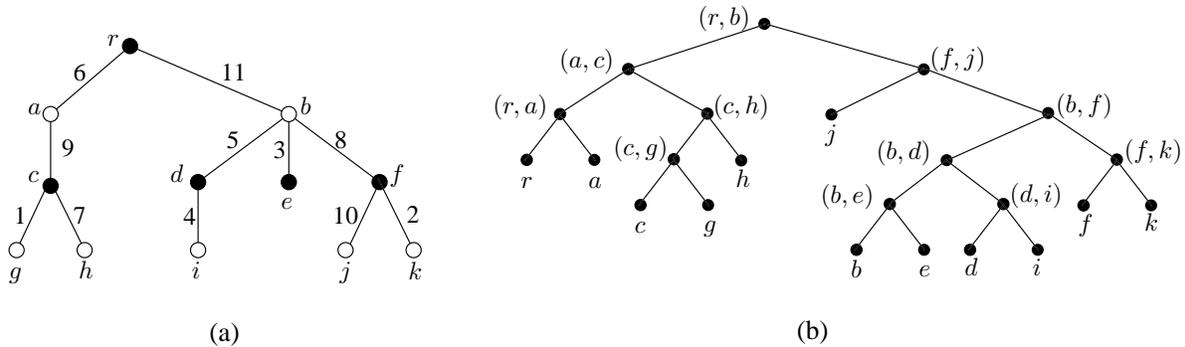}}
\end{center} \caption{
(a) The input weighted tree $T$; the filled nodes
are subtree roots when $T$ is partitioned with $g=3$.
(b) The Kruskal tree $K$ of $T$. Leaves
correspond to the nodes of $T$; internal nodes
correspond to edges of $T$.
\label{fig:kruskal}}
\end{figure}

The \boruvka~tree discussed in Section \ref{sec:mst} represents the
connected components that are formed as \boruvka's MST algorithm is
run.  We can define the analogous concept for other MST algorithms.
For example, the \emph{Kruskal tree} is the tree whose nodes are the
connected components formed as Kruskal's MST algorithm \cite{mst:kruskal} is run.
Kruskal's algorithm starts with all vertices in singleton components
and examines the edges in increasing order by weight, adding an edge
to the MST being built, and combining the two corresponding
components when the edge has ends in two different components.  The
Kruskal tree $K$ is binary, with one node per component, whose
children are the components combined to form the given component.
Each leaf of $K$ is a vertex of the original graph; each non-leaf
node is a non-singleton component. See Figure \ref{fig:kruskal}.

Even if the given graph is a tree, constructing the Kruskal tree is
equivalent to sorting the edges by weight, because the Kruskal tree
for a star (a tree of diameter two) contains enough information to
sort the edges.  If we are given the edges in order by weight,
however, the problem of constructing the Kruskal tree becomes more
interesting.  We shall develop an $O(n)$-time, pointer machine
algorithm to build the Kruskal tree $K$ of a tree $T$, given a list
of the edges of $T$ in order by weight.

\subsection{Bottom-Up Construction of a Kruskal Tree}
\label{sec:kt}

It is straightforward to build $K$ bottom-up using a DSU structure
whose nodes are the nodes of $T$ and whose sets are the node sets of
the current components.  As the algorithm proceeds, each designated
node of a set stores the node of $K$ corresponding to the set.  Root
$T$ at an arbitrary vertex; let $p(v)$ denote the parent of $v$ in
the rooted tree.  Initialize a DSU structure with each node in a
singleton set, storing itself (a leaf of $K$).  Process the edges
(now arcs) in the given order.  To process an arc $(p(v), v)$, let
$u = \find(p(v))$.  Add a new node $x$ to $K$, whose two children
are the nodes stored at $u$ and $v$.  Store $x$ at $u$, and perform
$\unite(u, v)$. (For example, in Figure \ref{fig:kruskal}, 
the node corresponding to $(f,j)$ is stored at $b$.)

This algorithm runs in $O(n\alpha(n, n))$ time on a pointer machine;
only the finds take non-linear time.  Although it builds $K$
bottom-up, it does not process $T$ bottom-up but in the given arc
order.  As in Sections \ref{sec:mst}--\ref{sec:dom}, 
we thus cannot directly apply the method
of Section \ref{sec:nca} to reduce the running time to linear.  On
the other hand, if we generalize the DSU structure to allow
$\mathit{unite}$ operations to have arbitrary nodes, rather than
just designated nodes, as parameters, and we replace each $\unite(u,
v)$ operation in the algorithm by $\unite(p(v), v)$, then the
(unordered) set of unions is known in advance, because the unions
correspond to the arcs of $T$.  As Thorup~\cite{usp:t99} observed
in the context of solving an equivalent problem (see Section
\ref{sec:ckt}), this means that the algorithm runs in linear time on
a RAM if the linear-time DSU algorithm of Gabow and Tarjan
\cite{dsu:gt} is used.

Not only are the unions not bottom-up on $T$, but also there is no obvious
way to transform the problem into one on a balanced tree as in
Section \ref{sec:mst}. Instead, we partition all of $T$ into
microtrees and do a topological graph computation to precompute the
answers to finds within the microtrees.  Once these answers are
known, running the algorithm to build $K$ takes $O(n)$ time.  Number
the arcs of $T$ from 1 through $n - 1$ in the given order.  For any
non-root vertex $v$, let $\num{v}$ be the number of $(p(v), v)$; let
$\num{v} = \infty$ if $v$ is the root. For any non-root vertex
$v$, let $f(v)$ be the node returned by $\find(p(v))$ in the
algorithm that builds $K$. (For example, in Figure \ref{fig:kruskal}, 
$f(j)=b$.)
Then $f(v)$ is the nearest ancestor $u$
of $v$ that has $\num{u} > \num{v}$. We will precompute $f(v)$ if
$v$ and $f(v)$ are in the same microtree.

\subsection{Linear-Time Construction}

Let $g = n/\log^{1/3}{n}$.  Partition all of $T$ into microtrees,
each of size at most $g$, using the method of Dixon, Rauch, and
Tarjan~\cite{mst:drt:j}, slightly modified. Visit the nodes of $T$ in a bottom-up
order, computing, for each node $v$, a size $s(v)$ and possibly
marking $v$ as a subtree root.  The value of $s(v)$ is the number of
descendants $w$ of $v$ such that no node on the path from $v$ to $w$
is marked.  When visiting $v$, set $s(v) \leftarrow 1 + \sum_{w\
\mbox{is a child of} \ v}{s(w)}$.  If $s(v) > g$, mark every child of
$v$ and set $s(v)$ to 1. Every marked node $v$ determines a
microtree whose nodes are the descendants $w$ of $v$ such that $v$
is the only marked node on the path from $v$ to $w$. The
construction guarantees that every microtree contains at most $g$
nodes.  It also guarantees that there are at most $n/g$ parents of
marked nodes, since, for each such parent, the set of microtrees
rooted at its children contains at least $g$ nodes. Partitioning $T$
into microtrees takes $O(n)$ time.

To precompute the answers to finds in the microtrees, begin by
initializing $f(v) \leftarrow \nul$ for every non-root node $v$. Then use a
pointer-based radix sort to renumber the nodes in each microtree
consecutively from 1 up to at most $g$ in an order consistent with
their original numbers (given by $\mathit{num}$).  This does not
affect the answers to the finds for any vertex whose answer is in
the same microtree.  To do the pointer-based radix sort, build a
master list of nodes representing the numbers 1 through $n$, and use
pointers to these nodes in lieu of the actual numbers.  For each
microtree, build a similar master list of nodes representing the
numbers 1 through the number of nodes in the microtree, and use
pointers to these nodes in lieu of numbers.  Now the problem of
answering the finds within microtrees is actually a topological
graph computation as defined in Section \ref{sec:tgc}, and with $g =
n/\log^{1/3}{n}$ it can be done in $O(n)$ time by Theorem \ref{thm:tgc2}.  
This computation
gives a non-null value $f(v)$ for every vertex $v$ such that $v$ and
$f(v)$ are in the same microtree.

Having precomputed the answers to some of the finds, we run the
algorithm that builds $K$, but using the precomputed answers.
Specifically, to process an arc $(p(v), v)$, let $u = f(v)$ if $f(v)
\neq \nul$, $u = \find(p(v))$ otherwise.  Then proceed as in Section
\ref{sec:kt}.

\begin{theorem}
Suppose that the edges of a weighted tree $T$
are given in order by weight. Then the Kruskal tree 
of $T$ can be built in $O(n)$ time on a pointer machine.
\end{theorem}
\begin{proof}
The algorithm runs on a pointer machine; the running time is $O(n)$
except for the time to do the finds.  We bound the time for the
finds by applying Lemma \ref{lemma:pc} to the tree built by the
parent assignments done by the unite operations.  Mark every parent
of a microtree root. This marks at most $n/g$ nodes.  If an
operation $\find(p(v))$ is actually done, because its answer is not
precomputed, $f(v)$ and $v$ are in different microtrees. The union
operations are such that if $x$ and $y$ are in the same set and $x$
is an ancestor of $y$, every vertex on the tree path from $x$ to $y$
is also in the same set. Thus when $\find(p(v))$ is done, $f(v)$,
$p(v)$, and $p(\mroot{\micro{v}})$ are all in the same set.  Since
$p(\mroot{\micro{v}})$ is marked, this find occurs in a set with a
marked node.  We conclude that Lemma \ref{lemma:pc} applies with $k
= 1$, giving an $O(n)$ time bound for the finds that are not
precomputed.
\end{proof}

We do not know whether there is a way to build $K$ in linear time
using only bottom-level microtrees.  If there is, it is likely to be
considerably more complicated than the algorithm we have proposed.

\subsection{Compressed Kruskal Trees}
\label{sec:ckt}

We can generalize the Kruskal tree to allow equal-weight edges: when
adding edges, we add all edges of the same weight at the same time
and add a node to the Kruskal tree for every new component so
formed, whose children are the components connected together to form
it.  The resulting component tree is not necessarily binary.
Thorup~\cite{usp:t99} and Pettie and Ramachandran~\cite{sp:pr02}
have used such a compressed Kruskal tree in shortest path
algorithms.  Given a tree and a partition of its edges into
equal-weight groups, ordered by weight, we can construct the
generalized Kruskal tree in linear time on a pointer machine as
follows.  Break ties in weight arbitrarily. Build the Kruskal tree,
labeling each component node with the group of the edge that formed
it.  Contract into a single node each connected set of nodes labeled
with the same group.  The last step is easy to do in $O(n)$ time.

\section{Concluding Remarks}
\label{sec:remarks}

We have presented linear-time pointer-machine algorithms for six
tree and graph problems, all of which have in common the need to
evaluate a function defined on paths in a tree.  Linear time is
optimal and matches the previous bound for RAM algorithms for
these problems; our algorithms improve previous pointer-machine
algorithms by an inverse-Ackermann-function factor.  Our
improvements rely mainly on three new ideas: refined analysis of
path compression when the compressions favor certain nodes;
pointer-based radix sort to help process small subproblems in batches;
and careful partitioning of the tree corresponding to the original
problem into a collection of microtrees and maximal paths, as
appropriate to the particular application.

\ignore{
An inverse-Ackermann-function factor is not noticeable in practice,
and we would not advocate actually using our algorithms, since the
previous slightly super-linear-time pointer machine algorithms are
simpler and almost certainly faster in practice.  Nevertheless, our
}
Our
algorithms are simpler than the previous linear-time RAM algorithms.
Indeed, our approach provides the first linear-time dominators algorithm that
could feasibly be implemented at all: the linear-time algorithm of
Alstrup et al.\cite{domin:ahlt99} requires Q-heaps \cite{mstj:fw},
implying an impossibly-large constant factor.  
Buchsbaum et al.~implemented their original RAM algorithm \cite{domin:bkrw},
of which
our pointer-machine algorithm is an improvement, and presented
experimental results demonstrating low constant factors, though the
simpler Lengauer-Tarjan algorithm was faster. 
Georgiadis, Tarjan, and Werneck
\cite{dom_exp:gtw06} report more recent experiments with algorithms for
finding dominators,
with results that vary depending on input size and complexity.

Our methods are sufficiently simple and general that we expect them
to have additional applications, which remain to be discovered.

\section*{Acknowledgements}

We thank 
Stephen Alstrup and Amos Fiat for 
some pointers
to previous works.

\bibliographystyle{plain}
\bibliography{compgeom,compilers,connect,deq,dstech,facts,mst,selfadj,skip,vlsi}

\end{document}